\documentclass[twocolumn,showpacs,prd]{revtex4}
\usepackage{graphicx}
\usepackage{dcolumn}
\usepackage{bm}
\begin{document}

\title{Ultra-High Energy Cosmic Ray Probes of Large Scale Structure
and Magnetic Fields\footnote{version published as  Phys. Rev. D 70,
043007 (2004), Copyright The American Physical Society 2004.}}
\author{G\"unter Sigl$^{a,b}$, Francesco Miniati$^c$,
Torsten~A.~En\ss lin$^c$}
\affiliation{$^a$ GReCO, Institut d'Astrophysique de Paris, C.N.R.S.,
98 bis boulevard Arago, F-75014 Paris, France\\
$^b$ F\'{e}d\'{e}ration de Recherche Astroparticule et Cosmologie, 
Universit\'{e} Paris 7, 2 place Jussieu, 75251 Paris Cedex 05, France \\
$^c$ Max-Planck Institut f\"ur Astrophysik,
Karl-Schwarzschild-Str.~1, 85741 Garching, Germany}

\begin{abstract}
We study signatures of a structured universe in the multi-pole
moments, auto-correlation function, and cluster statistics
of ultra-high energy cosmic rays above $10^{19}\,$eV. We compare
scenarios where the sources are distributed homogeneously or
according to the baryon density distribution obtained from
a cosmological large scale structure simulation. The influence
of extragalactic magnetic fields is studied by comparing
the case of negligible fields with fields expected to be produced
along large scale shocks with a maximal strength consistent with
observations. We confirm that strongly magnetized observers would
predict considerable anisotropy on large scales, which is already
in conflict with current data. In the best fit scenario only
the sources are strongly magnetized, although deflection
can still be considerable, of order $20^\circ$ up to $10^{20}\,$eV,
and a pronounced GZK cutoff is predicted.
We then discuss signatures for future large
scale full-sky detectors such as the Pierre Auger and EUSO projects.
Auto-correlations are sensitive to the source density only
if magnetic fields do not significantly affect propagation.
In contrast, for a weakly magnetized observer, degree scale
auto-correlations below a certain level indicate magnetized
discrete sources. It may be difficult even for next generation
experiments to distinguish between structured and unstructured
source distributions.
\end{abstract}

\pacs{98.70.Sa, 13.85.Tp, 98.65.Dx, 98.54.Cm}

\maketitle

\section{Introduction}

The origin of ultra-high energy cosmic rays (UHECRs) is still one of the most
challenging problems of modern astrophysics. It is an open question
which mechanism is responsible for producing particles up to $10^{20}\,$eV
and beyond and where the corresponding sources can be
found~\cite{reviews,school}.
Although statistically meaningful information about the UHECR energy
spectrum and arrival direction distribution has been accumulated, no
conclusive picture for the nature and distribution of the sources
emerges naturally from the data (for a short overview on the relevant
literature see Ref.~\cite{sme}). There is on the one hand the approximate
isotropic arrival direction distribution~\cite{bm} which indicates that we are
observing a large number of weak or distant sources. On the other hand,
there are also indications which point more towards a small number of
local and therefore bright sources: First, there seem to be
statistically significant multi-plets of events from the same directions
within a few degrees~\cite{bm,uchihori}. Second, nucleons above
$\simeq70\,$EeV suffer heavy energy losses due to
photo-pion production on the cosmic microwave background
--- the Greisen-Zatsepin-Kuzmin (GZK) effect~\cite{gzk} ---
which limits the distance to possible sources to less than
$\simeq100\,$Mpc~\cite{stecker}. For a uniform source distribution
this would predict a ``GZK cutoff'', a drop in the spectrum.
However, the existence of this ``cutoff'' is not established yet
from the observations~\cite{bergman}. In fact, whereas a cut-off
seems consistent
with the few events above $10^{20}\,$eV recorded by the fluorescence
detector HiRes~\cite{hires}, it is not compatible with the
11 events (also above $10^{20}\,$eV) measured by the AGASA ground
array~\cite{agasa}. The solution of this problem
may have to await the completion of the Pierre Auger project~\cite{auger}
which will combine the two complementary detection techniques
adopted by the aforementioned experiments.

Such apparently contradicting hints could easily be solved if it
would be possible to follow the UHECR trajectories backwards to their
sources. However, this may be complicated by the possible presence of
extragalactic magnetic fields, which would deflect the particles during
their travel. Furthermore, since the GZK-energy losses are of stochastic
nature, even a detailed knowledge of the extragalactic magnetic fields would
not necessarily allow to follow a UHECR trajectory backwards to its source
since the energy and therefor the Larmor radius of the particles
have changed in an
unknown way. Therefore it is not clear if charged particle astronomy with
UHECRs is possible in principle or not. And even if possible, it remains
unclear to which degree the angular 
resolution would be limited by magnetic deflection.

Quite a few simulations of the effect of extragalactic magnetic fields
(EGMF) on UHECRs exist in the literature, but usually idealizing
assumptions concerning properties and distributions of sources
or EGMF or both are made: In Refs.~\cite{slb,ils,lsb,sse,is} sources
and EGMF follow a pancake profile mimicking the local supergalactic
plane. In Ref.~\cite{dolag} highly structured EGMF have been obtained
from constrained simulations, but the source distribution has been
assumed homogeneous. In other studies EGMF have been approximated
in a number of fashions: as negligible~\cite{sommers,bdm},
as uniform~\cite{ynts}, or as organized in spatial cells
with a given coherence length and a strength
depending as a power law on the local density~\cite{tanco}.

However, the presence of the above mentioned {\it apparently contradicting
hints} indicate that the existing data set might already carry information
on non-trivial properties of sources and EGMF. Here, we want to address the
following questions relevant to charged particle astronomy:
\begin{enumerate}
  \item Do we observe a large number of dim or a small number of bright UHECR
  sources?
\item Is the source location distribution statistically homogeneous, or does
  it follow the matter distribution in the local Universe?
\item Are the particles strongly deflected by intergalactic magnetic fields?
\item What are the magnetic fields surrounding sources and observer ?
\item Can we discriminate between the case of sources with practical identical
  luminosities and the case of a power-law luminosity distribution?   
\end{enumerate}

The means by which we want to study these questions are comparisons of
simulated datasets to the observed one by statistical tests on the arrival
direction multi-pole moment and auto-correlation distributions, on the
multi-plet statistics, and on the UHECR energy spectrum.
The simulated UHECR events are produced by following the trajectories of
particles through a large-scale structure simulation which included a 
numerical model for the generation and evolution of the EGMF. We
thereby extend our former studies~\cite{sme,letter} 
to a larger parameter space and a higher degree of realism.

The most important results of these former studies are: A local component
of sources within $\lesssim100\,$Mpc alone cannot explain satisfactorily
the observed isotropy at energies $\lesssim4\times10^{19}\,$eV~\cite{sme}.
In Ref.~\cite{letter} and the present study we therefore take into
account sources at cosmological distances by periodically repeating
the large scale structure simulation box. As a result we
found that in combination, a comparison of spherical multi-poles for $l\leq10$
and of the auto-correlation at angles $\theta\lesssim20^\circ$ between
observed and simulated data moderately favors a scenario in which (i) UHECR
sources have a density $n_s\sim10^{-5}\,{\rm Mpc}^{-3}$ and follow the matter
distribution (ii) magnetic fields are relatively pervasive within the large
scale structure, including filaments, and with a strength of order of a $\mu$G
in galaxy clusters (iii) the local extragalactic environment is characterized
by a weak magnetic field below $ 0.1\,\mu$G. This is in contrast to
Ref.~\cite{sme} where the neglect of cosmological sources marginally
favored observers immersed in $\sim0.1\,\mu$G fields. Finally, we
found that the degree-scale
auto-correlation functions above $\simeq4\times10^{19}\,$eV can serve as a
discriminator between magnetized and unmagnetized sources.

In the present paper we specifically focus on signatures for the structure,
density and luminosity distribution of UHECR sources, as well
as for magnetic fields surrounding sources and observer. We will
find, as expected, that for most of these observables, the current
data set does not allow to clearly distinguish between limiting
cases. We will therefore discuss how future experiments such as
the Pierre Auger~\cite{auger} and EUSO~\cite{euso} projects will
improve the prospects to measure these observables.

Here, we restrict ourselves to UHECR nucleons, and we neglect the
Galactic contribution to the deflection of UHECR since
typical proton deflection angles in galactic magnetic fields of
several $\mu$G are $\lesssim10^\circ$ above
$4\times10^{19}\,$eV~\cite{medina}, and thus in general are small
compared to extragalactic deflection in the scenarios studied
in the present paper.

The simulations are described in more detail in the next section.
There we also describe the general features of our method
and define the statistical quantities used for comparison with
the data. Sect.~3 presents how large scale multi-poles and
small-scale auto-correlations probe magnetization and UHECR source
characteristics. It constitutes
the main part of the present paper. We conclude in Sect.~4.

\section{Outline of the Numerical Technique}

In the next section we will investigate, from a statistical point 
of view as defined below, the viability of various scenarios for the 
propagation of UHECRs in an extragalactic environment. These scenarios
are listed in Tab.~\ref{tab1} and differ in the UHECR source distributions,
the strength of the EGMF, and the location of an hypothetical
observer on Earth. The latter will also lead to different
strengths of the EGMF within a few Mpc from the observer.
In the following sections we describe in detail how the various
scenarios are characterized.

\subsection{Large Scale Structure and Extragalactic Magnetic Fields}

The magnetized extragalactic environment which we use for our experiments
is produced by a simulation of the large scale structure of
the Universe. The simulation was carried out
within a computational box of $50\,h^{-1}\,$Mpc length on a side, 
with normalized Hubble constant 
$h\equiv H_0/(100$ km s$^{-1}$ Mpc$^{-1})$ = 0.67, and using
a comoving grid of 512$^3$ zones and 256$^3$ dark matter
particles. This is the same large scale structure simulation that was used in
Refs.~\cite{sme,letter} and is further described in~\cite{miniati}.

\begin{figure}[ht]
\includegraphics[width=0.45\textwidth,clip=true]{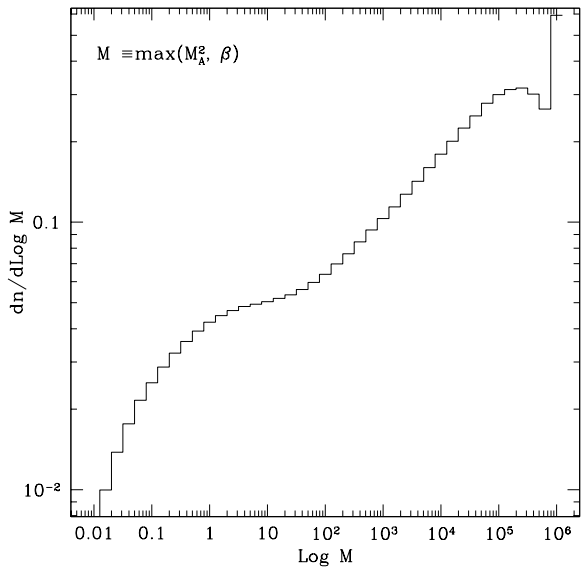}
\caption[...]{Distribution
of the number of cells n as a function of the maximum M of
the Alfv\`en Mach number and the plasma beta parameter.}
\label{fig0}
\end{figure}

The EGMF adopted here is based on the numerical model developed in 
\cite{ryu} which has been shown to be compatible with existing 
Faraday rotation measures with lines of sight both through clusters and
the diffuse intergalactic medium. Thus, at simulation start the EGMF
was initialized to zero and subsequently its seeds were generated at
cosmic shocks through the Biermann battery
mechanism~\cite{kcor97}. This approach is alternative to the case in
which the initial magnetic field is set uniform over the whole
simulated volume. Since cosmic shocks form primarily around collapsing
structures including filaments, the above approach avoids generating
EGMF in cosmic voids. An alternative, more realistic but also much
more complicated scenario is discussed in Ref.~ \cite{kronberg99}, in
which magnetic fields are injected into the intergalactic medium by
galactic outflows.  Whichever the mechanism that generates it, the
magnetic field is then evolved according to the induction equation and
is therefore amplified in different parts of the universe by shear
flows and compression according to the velocity field provided by the
simulated gas component.

\begin{figure}[ht]
\includegraphics[width=0.49\textwidth,clip=true]{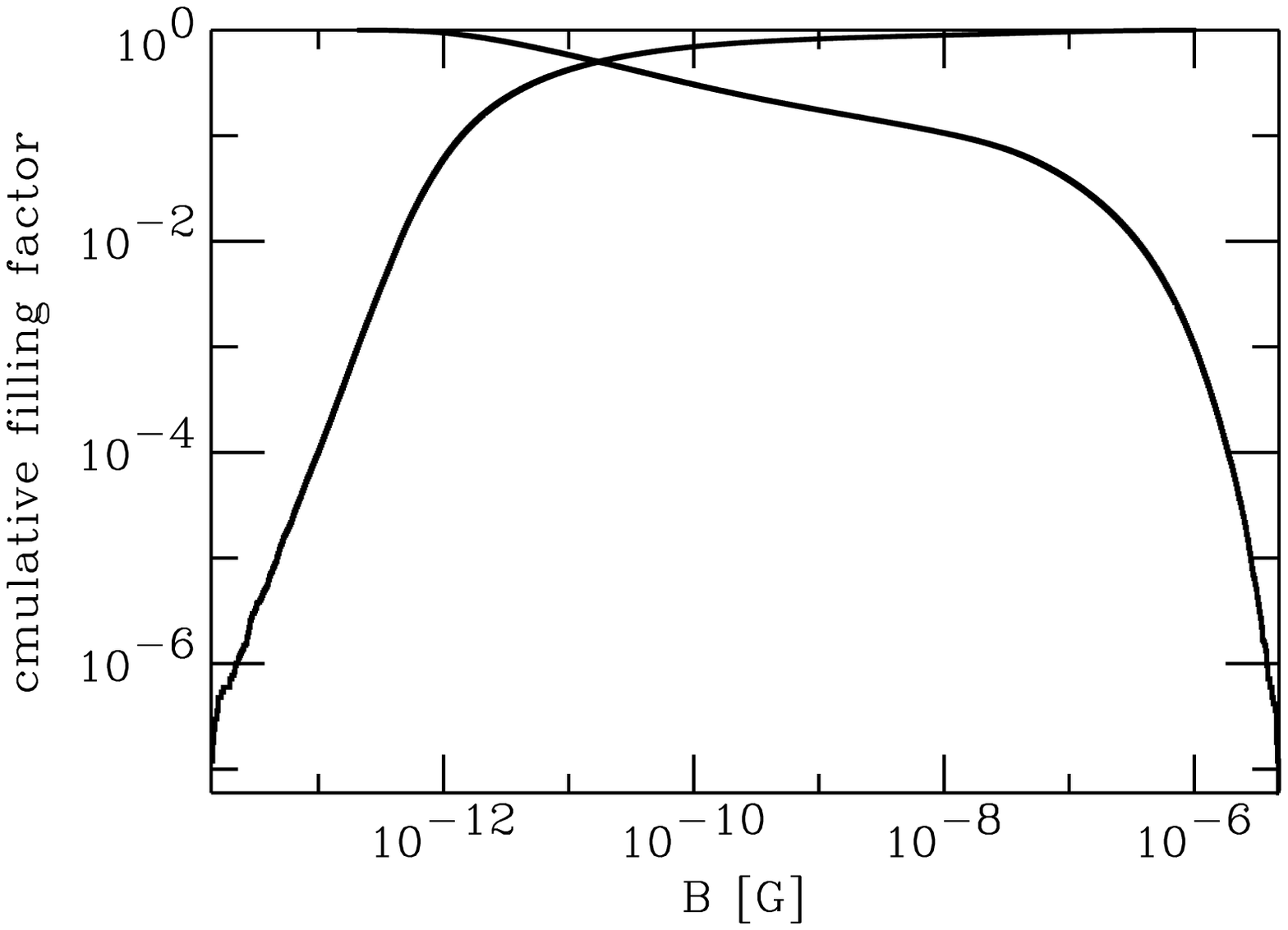}
\includegraphics[width=0.49\textwidth,clip=true]{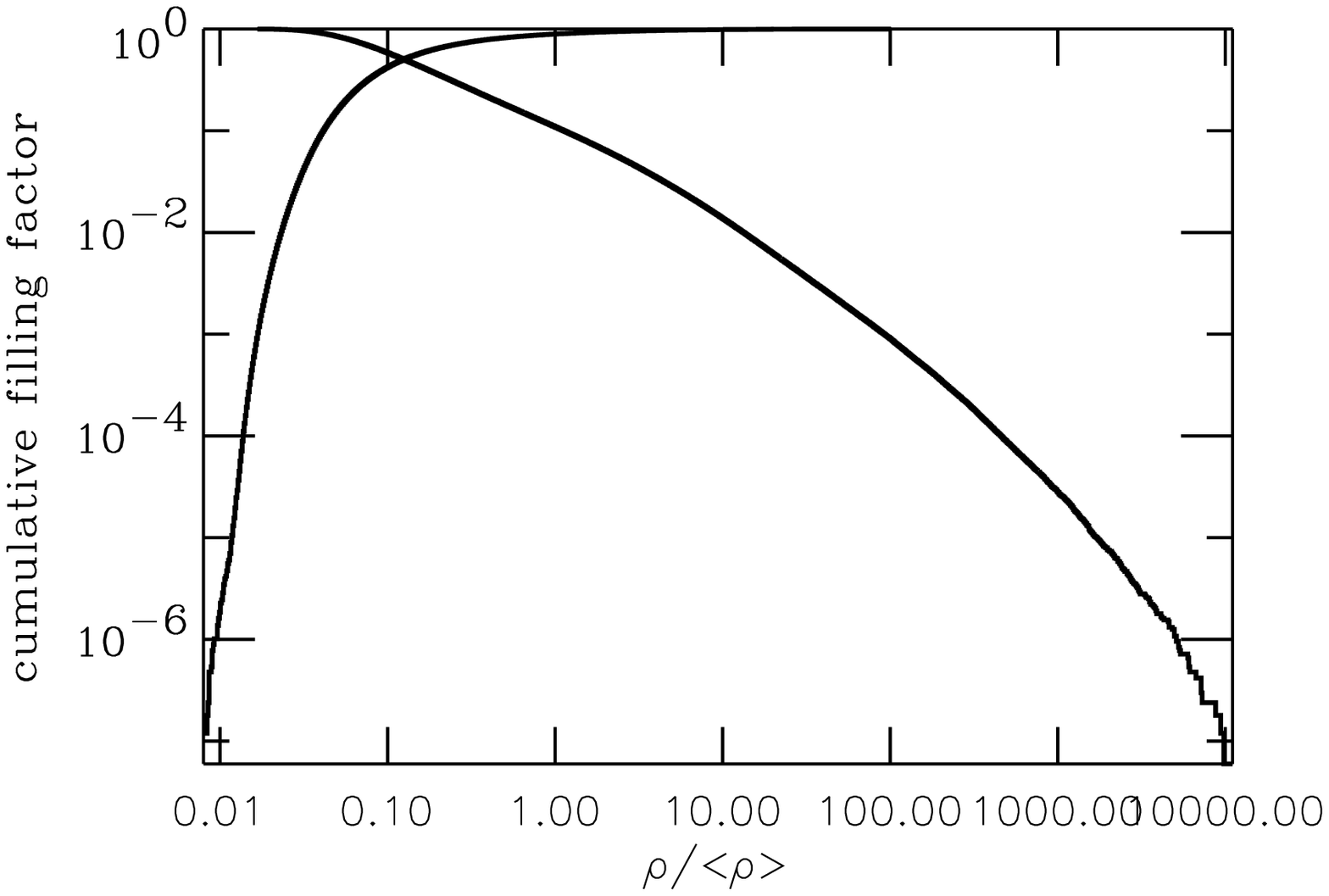}
\caption[...]{The cumulative filling factors for EGMF strength,
(middle panel) baryon density (lower panel, in units of average
baryon density) above (decreasing curves) and below (increasing curves)
a given threshold, as a function of that threshold.}
\label{fig1}
\end{figure}

As already pointed out in Ref.~\cite{sme}, given the tiny values of
the initial seeds generated through the Biermann mechanism, the field
strength at simulation end (when the cosmological redshift equals
zero) is much smaller than what is observed in galaxy clusters via a
number of experiments~\cite{bo_review}.  Therefore, in order to avoid
such discrepancies, it is necessary to change the normalization of the
simulated magnetic field strength. Our renormalization procedure
simply involves a rescaling of the overall magnetic field in the
computational box, such that the magnetic field in the core region of
a Coma-like galaxy cluster is {\it predicted} to be of order of a
$\mu$G or so, as indicated by Faraday rotation measures
\cite{bo_review}.  As a result of this rescaling, the magnetic field
strength volume averaged over $\simeq0.5\,$Mpc within typical cluster
cores is between 0.7 and 2.5 $\mu$G.  Since the magnetic field
strength, $B$, in collapsed structures follows a well defined scaling
relation with the structures virial temperature~\cite{minthe}, the
renormalization of the magnetic field as described above can be easily
carried out even though Coma like galaxy clusters do not form in our
simulation due to the relatively small computational box.

Lacking direct measurements of magnetic fields in filaments, we assume
that the topology and relative strength of the large scale
intergalactic magnetic field in different parts of the Universe is as
reproduced by our numerical simulation.  However, as already pointed out 
the resulting EGMF is consistent with statistics of existing 
Faraday rotation measures with line of sight through filaments despite
the fact that the magnetic field strenght can be close to the equipartition 
value with the gas total energy \cite{ryu}.

\begin{figure}[ht]
\includegraphics[width=0.49\textwidth,clip=true]{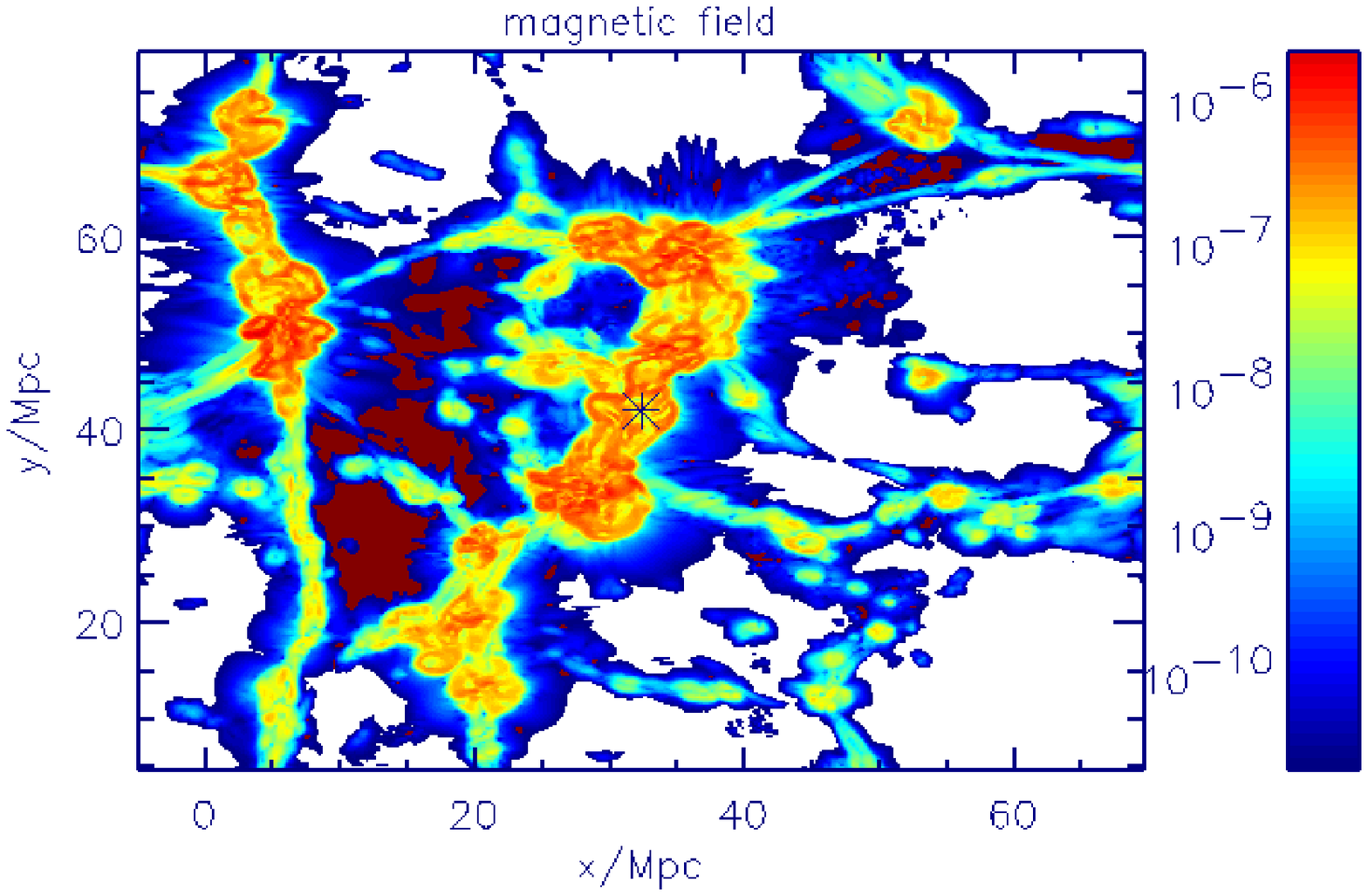}
\includegraphics[width=0.49\textwidth,clip=true]{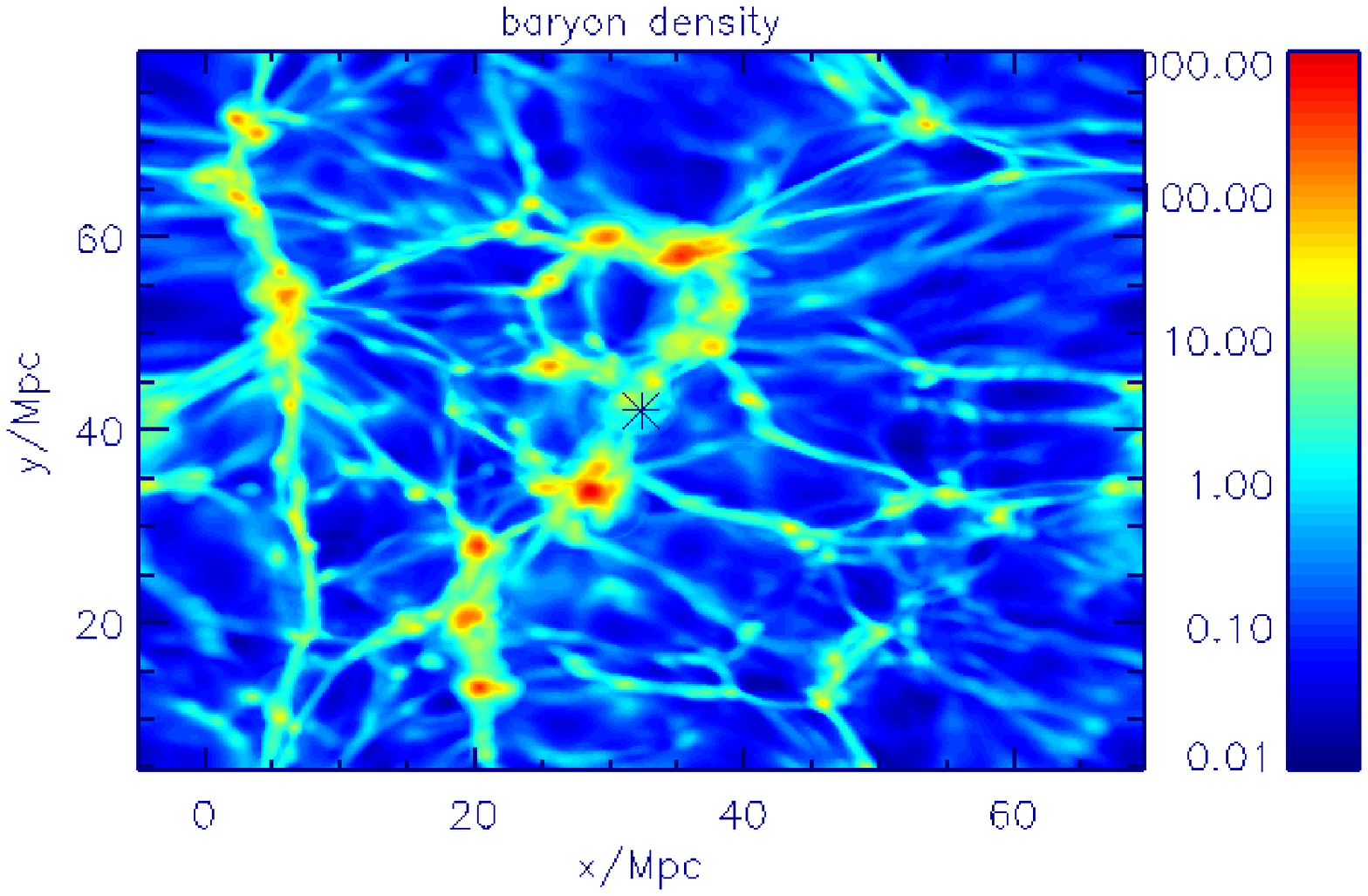}
\caption[...]{Log-scale two-dimensional cut through
magnetic field total strength in Gauss (color
scale in Gauss, upper panel) and baryon density in units of average
baryon density (color scale, lower panel).
The observer is in the center of the figures and is marked by a star.
The EGMF strength at the observer is $\simeq0.1\,\mu$G. Note that
both panels correspond to the same cuts through the full large scale
simulation box.}
\label{fig2}
\end{figure}

\begin{figure}[ht]
\includegraphics[width=0.49\textwidth,clip=true]{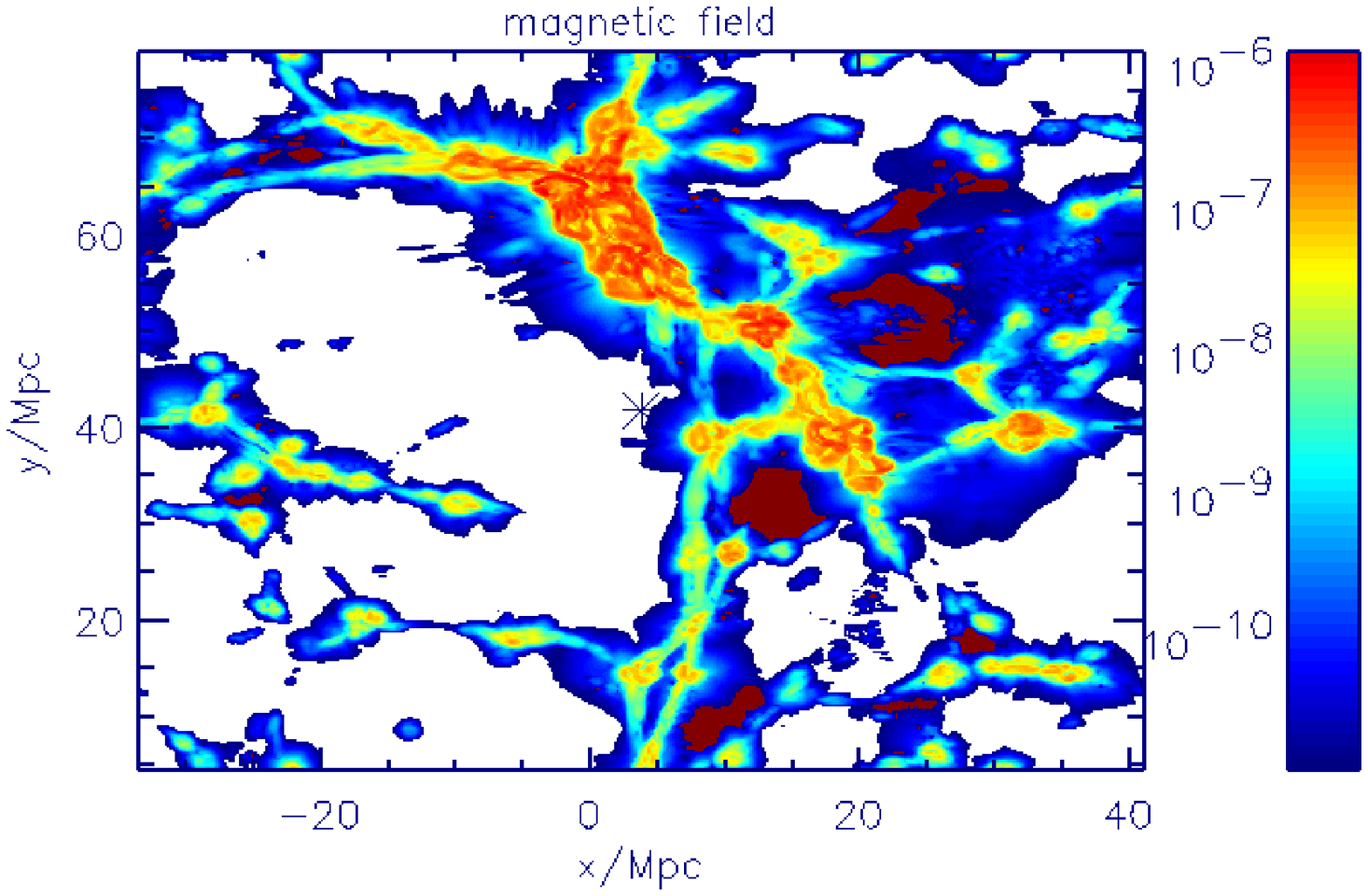}
\includegraphics[width=0.49\textwidth,clip=true]{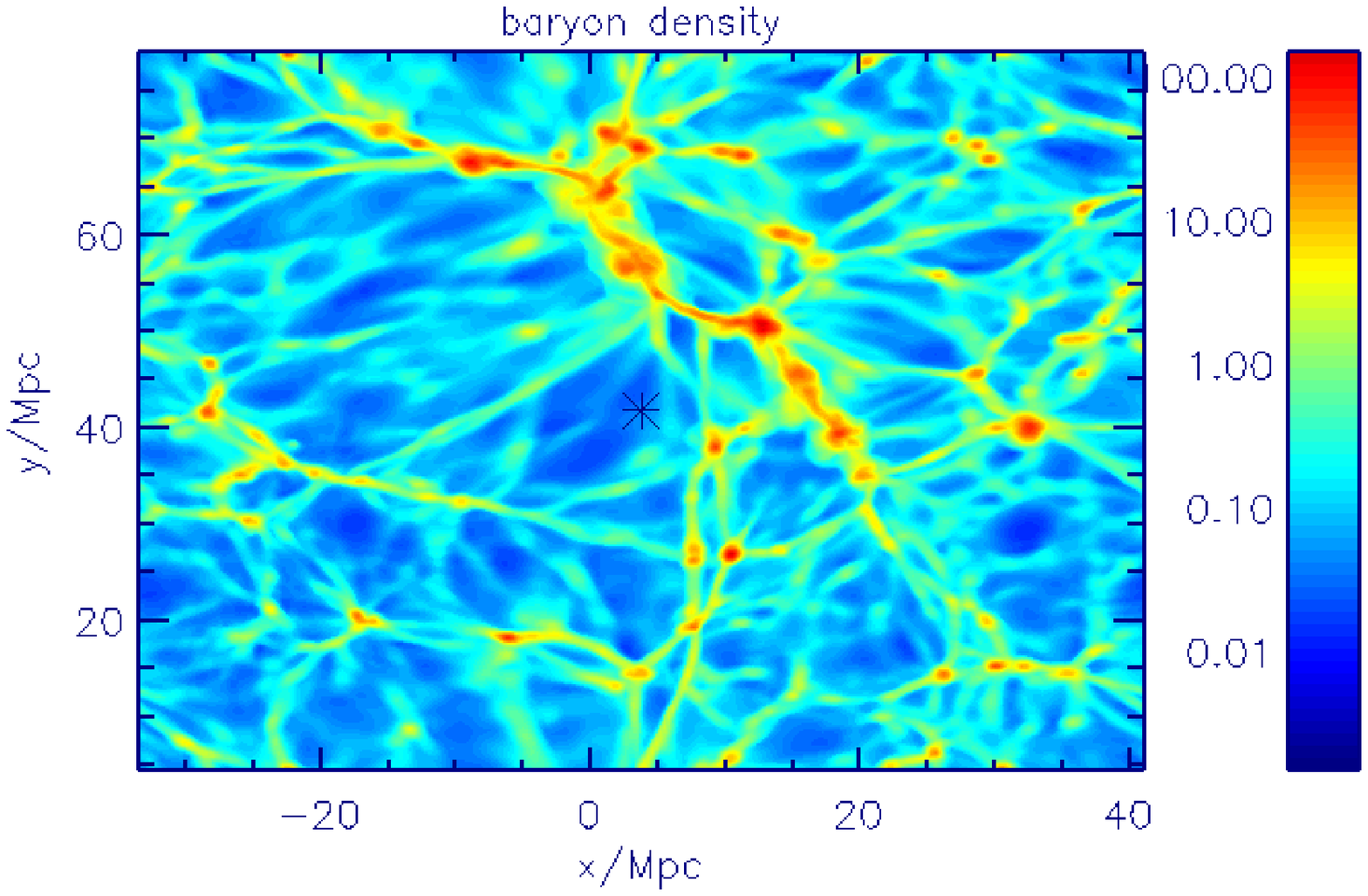}
\caption[...]{Same as Fig.~\ref{fig2}, but for an observer
situated in a small void where the EGMF strength is $\simeq10^{-11}\,$G.}
\label{fig3}
\end{figure}

It is worth pointing out that, because of the very scarce observational
constraints on intergalactic magnetic fields, both our assumptions
have limitations. For example, an independent experiment based on the
detection from the Coma cluster of radio synchrotron emission and hard
X-rays interpreted as inverse Compton emission, would suggest that
magnetic fields in this cluster are lower by about an order of magnitude
compared to what we are assuming~\cite{fufe99}. Similarly, and more
importantly, one cannot exclude
the possibility of much smaller magnetic fields in filaments than 
we assume, although some evidence for magnetic fields in filaments
at the level of a tenth of a $\mu$G may already exist~\cite{bagchi02}.

According to our simulation scenario and with the assumptions made, we
find that EGMF are significant only within filaments and
groups/clusters of galaxies. In Fig.~\ref{fig0} we test the assumption
that the magnetic field is passive. In order for the magnetic field to
be dynamically important both the Alfv\`enic Mach number, $M_A=v/v_A=
v/B/(4\pi\rho)^{1/2}$, and the plasma $\beta$ parameter, $\beta=
P_{gas}/(B^2/8\pi)$ must be smaller than unity. Thus in
Fig.~\ref{fig0} we plot a histogram of the fraction of cells as a
function of $M=\max(M_A^2, \beta)$. The histogram shows that the
condition of dynamically unimportant magnetic field is violated only
in a very small fraction of the volume, which does not affect the
evolution of the simulation in any significant way. 
Furthermore, only a fraction of the cells
in those bins is characterized by a a magnetic field capable of
affecting the trajectory of UHECRs. In any case, as discussed in
Sec. \ref{disc.sec} we consider the case of a field strength
normalization reduded by a factor 10 with respect to our fiducial
model. This would correspond to a left-shift of the x-axis in
Fig.~\ref{fig0} of two decades, such that the EGMF is virtually
dynamically unimportant in all cells of the simulation.

About 90 percent of the volume is filled with fields $\lesssim10\,$nG
and in the voids fields are $\lesssim10^{-11}\,$G. In Fig.~\ref{fig1}
we show the cumulative filling factors for both the EGMF strength and
the baryonic gas. The volume field averages are
$\left\langle |B|\right\rangle\simeq1.5\times10^{-8}\,$G, and
$\left(\left\langle B^2\right\rangle\right)^{1/2}\simeq7.9\times10^{-8}\,$G,
with coherence lengths $\lesssim\,$Mpc in the strong field regions, and thus
compatible with Faraday rotation bounds~\cite{bo_review,ryu}.
Note that the hallmark of a highly structured field is a ratio
$\left(\left\langle B^2\right\rangle\right)^{1/2}/
\left\langle |B|\right\rangle\gg1$, as in the present case.

As in Refs.~\cite{sme,letter}, we explore the case of two observers
located at different positions within the simulated volume: 
The first is located in a filament-like structure with EGMF
$\sim0.1\,\mu$G, and the second at the border of a small void with
EGMF $\sim10^{-11}\,$G. 
Figs.~\ref{fig2} and~\ref{fig3} show two-dimensional cuts through
the EGMF and baryon density around these two observer positions.
Notice that the structure of the magnetic field is quite more extended
than that for the baryonic gas. Thus a UHECR produced by a source where
matter density is high will be subject to the action of magnetic fields
within an extended volume surrounding the source, before breaking into 
a void where magnetic fields are much weaker.
A relatively large structure about 17 Mpc away
from the weak field observer is identified for calculation purposes 
as the Virgo cluster. We orient our terrestrial coordinate system so
that this cluster is close to the equatorial plane.

\subsection{UHECR Sources}

For a given number density of UHECR sources, $n_s$, 
we explore both the case in which their spatial distribution is 
either proportional to the local baryon density, as in Ref.~\cite{sme}, or
completely homogeneous. Further, due to the unknown source positions
and properties, there will be a cosmic variance in the results.
In order to be conservative and maximize this variance, we will
therefore assume that all UHECR sources, are distributed in
luminosity, $Q_i$, so that their contribution of UHECR per $\log Q_i$
is roughly constant with $Q_i$. In addition we allow the spectral index 
$\alpha_i$ of the emitted power-law
distributions of UHECRs to vary and assume that each 
source accelerates UHECRs up to $10^{21}\,$eV. 
Specifically, our assumption can be summarized as follows,
\begin{eqnarray}
  \frac{dn_s}{dQ_i}&\propto&Q_i^{-2.2}\quad\mbox{for}\,\,
  1\leq Q_i\leq Q_{\rm max}  \nonumber\\
  \frac{dn_s}{d\alpha_i}&=&{\rm const}\quad\mbox{for}\,\,
  \alpha-\Delta\alpha\leq\alpha_i\leq\alpha+\Delta\alpha \label{source_prop}\\
  E_{\rm max}&=&10^{21}\,{\rm eV}\,,\nonumber
\end{eqnarray}
where $Q_i$ has been put in dimensionless units. The power law distribution
in $Q_i$ could be further
motivated by the luminosity function of the EGRET $\gamma-$ray
blazars which has this shape in the power range
$10^{46}\,{\rm erg}\,{\rm s}^{-1}\lesssim Q_i
\lesssim10^{48}\,{\rm erg}\,{\rm s}^{-1}$~\cite{cm}.
Constant source characteristics correspond to $Q_{\rm max}=1$,
$\Delta\alpha=0$ in Eq.~(\ref{source_prop}). If not explicitly
stated that sources with identical properties are assumed,
we will always use $Q_{\rm max}=100$, $\Delta\alpha=0.1$.
Finally, the actual value of $\alpha$, representing the central value
of the power law index distribution, together with the total power of
injected UHECRs are left as free parameters to be obtained from a best
fit analysis when the simulation results are compared with
observational data.

\begin{figure}[ht]
\includegraphics[width=0.49\textwidth,clip=true]{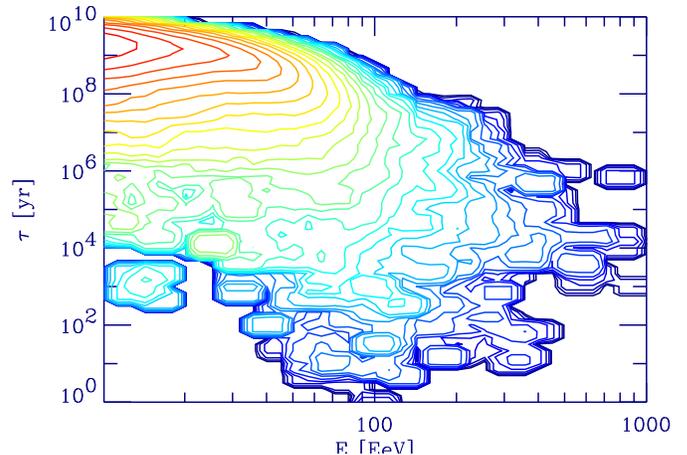}
\caption[...]{The distribution of UHECR detection energy $E$ and
delay time $\tau$ with respect to straight line propagation,
for scenario 6 in Tab.~\ref{tab1}, averaged over 26 realizations
of $10^4$ simulated trajectories above $10^{19}\,$eV each.}
\label{fig4}
\end{figure}

We also assume that neither total power, $Q_i$,
nor the power law spectral index, $\alpha_i$,
change significantly on the time scale of UHECR propagation. 
For energies higher than the GZK cutoff this is not an issue because the 
sources are nearby and the propagation time, given by the distance in 
light years added by the time delay provided in Fig.~\ref{fig4},
is less than the typical duty cycle of, say, a radio source. 
On the other hand, for UHECRs of lower energies the propagation time 
can be up to a few Gyr, see Fig.~\ref{fig4}. However, the flux of UHECRs
at these energies is dominated by many sources at relatively large distances
and, therefore, should not be sensitive to time variations of individual
sources.

Taking advantage of the periodic boundary conditions, we build a
``hyper volume'' by adding 
periodic images of the simulation box until the linear
size of the enclosed volume is $\sim 1.5 h^{-1}$ Gpc, that is 
larger than the energy loss length of nucleons above $10^{19}\,$eV.
The distributions of sources and the EGMF are identical
in each box for a given realization. Further, in each replicated 
box within the central Gpc of the hyper volume we place
an observer at the same position (relative to the box).
This ensures that each observer is still surrounded by several
100 Mpc of sources in each direction. Since the energy loss length
for nucleons is $\lesssim1\,$Gpc above $10^{19}\,$eV~\cite{bs-rev},
all relevant distant sources are taken into account in this manner.

\subsection{UHECR Propagation and Event Detection}

For each configuration, consisting of a choice of source positions 
and a set of parameters $Q_i$ and $\alpha_i$, 
many nucleon trajectories originating from 
sources in the Gpc$^3$  volume 
were computed numerically.
As in previous work ~\cite{slb,ils,lsb,is,sme}, particles are propagated
taking into account Lorentz forces due to the EGMF and energy losses.
In this respect, pion production is treated stochastically,
whereas pair production is treated as a continuous energy loss process.
Cosmological
redshift cannot be taken into account because the propagation time
is not known before hand. However, this is a minor effect
since the travel time of the trajectories does not
exceed $\sim3\,$Gyr, corresponding to $z\sim0.6$ down to $10^{19}\,$eV,
see Fig.~\ref{fig4}.
Trajectories connecting sources and observers in different copies
of the simulation box are taken into account.

An {\it event} was registered and its arrival direction and 
energy recorded each time the trajectory of the propagating particle
crossed a sphere of radius 1.5 Mpc around one of the observers.
Events at replicated observers in different boxes are recorded by the
same counter. The counter is stopped when $10^4$ events
(a realization) have been recorded.
For more details on this method see Refs.~\cite{slb,lsb,ils}.
Modeling the observer as a sphere corresponds to an average over
observers located on that sphere. The concrete directions of features
in the UHECR sky distribution can never be accurately reconstructed
in this approach. However, quantities which only depend on angular
distances such as the multi-poles, the auto-correlation function,
and the number of event clusters can properly be accounted for by
this approach.

Having provided all the relevant definitions, for various scenarios
explored in the next section we build statistical samples in the
following way.
We generate ten realizations for the source positions. For each one
of them we use the distributions in Eq.~(\ref{source_prop}),
with $Q_{\rm max}=100$, $\Delta\alpha=0.1$, and
typically generate 50 realizations with different power $Q_i$ and
injection spectral index $\alpha_i$, whereas the maximal acceleration
energy is held constant at $10^{21}\,$eV for simplicity. Thus,
for each scenario typically 500 realizations corresponding to
different configurations of source positions and
emission characteristics are simulated. The variation
of results with these configurations constitutes cosmic variance.

\subsection{Data Processing}

The process of converting the simulated events to quantities that can
be compared with observations was described at length in Ref.~\cite{sme}.
Here we simply summarize the procedure.

For each realization discussed above, the events were used
to construct arrival direction probability distributions
above a given energy on regular grids of solid angle bins of size
$\Delta\theta$ in both declination and right ascension. For the full sky,
for example, there are $\simeq180^\circ/\Delta\theta$ bins in declination
and, at declination $\delta$, $\simeq(360^\circ/\Delta\theta)\cos\delta$
bins in right ascension. We chose $\Delta\theta$ small compared to
the instrumental angular resolution; typically $\Delta\theta=0.5^\circ$.
The sky distributions are multiplied with the solid-angle dependent
exposure function for the respective experiment and are convolved with
the angular resolution which is $\simeq1.6^\circ$ above
$4\times10^{19}\,$eV and $\simeq2.5^\circ$ above $10^{19}\,$eV for the AGASA
experiment~\cite{agasa}. Due to the poorer angular resolution of the
SUGAR data~\cite{sugar}, we will use a resolution of $2.5^\circ$ when combining
AGASA and SUGAR data. Energy resolution effects, supposed to be of
order $\Delta E/E\simeq30\%$, are also taken into account.

For the exposure function $\omega(\delta)$ we use the parameterization
of Ref.~\cite{sommers} which depends only on declination $\delta$,
\begin{eqnarray}
\omega(\delta)&\propto&\cos(a_0)\cos(\delta)\sin(\alpha_m)+
        \alpha_m\sin(a_0)\sin(\delta)\,,\nonumber\\
&&\mbox{where}\hskip0.5cm\alpha_m = \left\{ \begin{array}{ll}
\ 0 & \mbox{if $\xi > 1$}\\
\ \pi & \mbox{if $\xi < -1$}\\
\ \cos^{-1}(\xi) & \mbox{otherwise}
 \end{array} \right.\,,\label{expos}\\
&&\mbox{with}\hskip0.5cm\xi\equiv
\frac{\cos(\theta_m)-\sin(a_0)\sin(\delta)}{\cos(a_0)\cos(\delta)}\,.
\nonumber
\end{eqnarray}
For the AGASA experiment~\cite{agasa} $a_0=35^\circ$, $\theta_m=60^\circ$,
and for the former SUGAR experiment~\cite{sugar} on the southern hemisphere
$a_0=-30.5^\circ$, $\theta_m=55^\circ$. For a full-sky Pierre
Auger type experiment we add the exposures for the Southern Auger
site with $a_0=-35^\circ$ and a putative similar Northern site with
$a_0=39^\circ$, and $\theta_m=60^\circ$ in both cases,
with an assumed angular resolution of $\simeq1^\circ$.

From the distributions obtained in this way typically 200 mock data sets
consisting of $N_{\rm obs}$ observed events were selected randomly.
For each such mock data set or for the real data
set we then obtained estimators for the spherical harmonic coefficients
$C(l)$, the auto-correlation function $N(\theta)$, and the number
of multi-plets $M(n)$ of $n$ events within an angle $\theta_m$.
As in Refs.~\cite{sme,letter} the estimator for $C(l)$ is defined as
\begin{equation}
  C(l)=\frac{1}{2l+1}\frac{1}{{\cal N}^2}
  \sum_{m=-l}^l\left(\sum_{i=1}^{N_{\rm obs}}\frac{1}{\omega_i}Y_{lm}(u^i)
  \right)^2\,,\label{cl}
\end{equation}
where $\omega_i$ is the total experimental exposure
at arrival direction $u^i$, ${\cal N}=\sum_{i=1}^{N_{\rm obs}}1/\omega_i$
is the sum of the weights $1/\omega_i$, and
$Y_{lm}(u^i)$ is the real-valued spherical harmonics function
taken at direction $u^i$. Also as in Refs.~\cite{sme,letter} the
estimator for $N(\theta)$ is defined as
\begin{equation}
N(\theta)=\frac{C}{S(\theta)}\sum_{j \neq i}
\left\{\begin{array}{ll}
1 & \mbox {if $\theta_{ij}$ is in same bin as $\theta$}\\
0 & \mbox{otherwise}
\end{array}\right\}\,,
\label{auto}
\end{equation}
and $S(\theta)$ is the solid angle size of the corresponding bin.
In Eq.~(\ref{auto}) the normalization factor
$C=\Omega_e/(N_{\rm obs}(N_{\rm obs}-1))$,
with $\Omega_e$ denoting the solid angle of the sky region where the
experiment has non-vanishing exposure, is chosen such that an
isotropic distribution corresponds to $N(\theta)=1$.

Finally, the multi-plets are obtained as follows: For a given set of
$N_{\rm obs}$ ordered events, the first event is considered as a
singlet. Then for all other events we check if they lie within distance
$\theta_m$ of one of the other events defining the center of a multi-plet.
If yes, the multiplicity of that multi-plet is increased by one, if
not, the event is defined as the center of a new multi-plet, starting
as a singlet.

In passing we note that other statistical quantities have been
considered in the literature, such as cumulative distributions
in right ascension and declination in the context of the Kolmogorov-Smirnov
test~\cite{k-s}, and the ``information-dimension'' of the sky
distribution~\cite{sjm}. We will not use these statistics in the
present work.

The different mock data sets in the various realizations
yield the statistical distributions of $C(l)$, $N(\theta)$, and
$M(n)$. One defines the average over all mock data sets and realizations
as well as two errors. The smaller error (shown to the left of
the average in the figures below) is the statistical error, i.e. the
fluctuations due to the finite number $N_{\rm obs}$ of observed
events, averaged over all realizations. The larger error
(shown to the right of the average in the figures below)
is the ``total error'', i.e. the statistical error plus the
cosmic variance.
Thus, the latter includes the fluctuations due to finite
number of events and the variation between different realizations
of observer and source positions.

Given a set of observed and simulated events, after extracting
some useful statistical quantities $S_i$,
namely $C_l$, $N(\theta)$, and $M(n)$ defined above, we define
\begin{equation}
  \chi_n\equiv\sum_i
  \left(\frac{S_{i,{\rm data}}-\overline{S}_{i,{\rm simu}}}
             {\Delta S_{i,{\rm simu}}}\right)^n\, \label{chi_n}
\end{equation}
where the index $i$ runs over multi-pole $l$, angular bin of $\theta$,
and multiplicity $n$, respectively.
Here, $S_{i,{\rm data}}$ refers to $S_i$ obtained from either
the real data set or the simulated mock data sets, and
$\overline{S}_{i,{\rm simu}}$ and $\Delta S_{i,{\rm simu}}$ are the
average and standard deviations of the simulated data sets. Thus,
there is a $\chi_n$ for the real data, and a $\chi_n$ for each of
the simulated mock data sets which consist of the same total
number of UHECR events. This measure of deviation from the average
prediction can then be used to obtain an overall likelihood for
the consistency of a given theoretical model with an observed data
set by counting the fraction of simulated data sets with
$\chi_n$ larger than the one for the real data.

\section{Results: Probing EGMF and UHECR Source Characteristics}

We now turn to a systematic discussion of signatures of magnetization
and UHECR source characteristics in the angular power spectrum,
the auto-correlation function, and the clustering of the UHECR arrival
distributions. 

The scenarios studied are presented in Tab.~\ref{tab1} together with 
a statistical measure of their likelihoods. To summarize,
UHECR sources whose number density is given in column 2
are distributed either proportionally to the simulated baryonic
density or homogeneously (``yes'' or ``no'', respectively,
in column 3). The observer is either in a region with or 
without appreciable magnetic fields, as quantified more precisely 
in Fig.~\ref{fig2} and~\ref{fig3}, respectively. 
Finally, the EGMF is either taken from the simulation with local value
as indicated in column 5, or completely neglected (``no EGMF'').
For all scenarios our statistical assessments are based on
typically 200 simulated mock data sets each for 500 different
realizations of source locations and emission characteristics,
as discussed in Sects.II.C and II.D above.
Scenarios 1-6 in Tab.~\ref{tab1} have already
been presented in Ref.~\cite{letter}, whereas scenarios 7-10 are new.

\begin{table*}[ht]
\caption[...]{List of simulated scenarios. The columns contain
the number assigned to the scenario, the source density, whether the
sources are distributed as the baryon density in the large scale
structure simulation box or homogeneously (yes/no), the observer position
1 or 2 corresponding to Fig.~\ref{fig2}
and~\ref{fig3}, respectively, the magnetic field
strength at the observer location (zero indicates no fields, whereas
a number indicates the EGMF obtained from the large scale structure
simulation), the best fit power law index
in the injection spectrum $E^{-\alpha}$, and the overall likelihoods
of fits to the data. The first six likelihoods are for the
multi-poles Eq.~(\ref{cl}) above the energy indicated as superscript
in EeV and over the range of $l$ indicated as subscript. ``AGASA only''
and ``AGASA+SUGAR'' indicates which exposure functions and data sets
were used. Above 40 EeV this corresponds to $N_{\rm obs}=99$
``AGASA+SUGAR'' events, or to $N_{\rm obs}=57$ ``AGASA only'' events.
Above 10 EeV comparison with an isotropic distribution of 1500 events
was made, see text for more details.
The last two likelihoods are for the auto-correlation
Eq.~(\ref{auto}) for $\theta\leq20^\circ$, and the clustering within
$2.5^\circ$ up to multiplicity 10, respectively. The
likelihoods are computed for $n=4$ in Eq.~(\ref{chi_n}) which
leads to reasonable discriminative power.}\label{tab1}
\begin{ruledtabular}
\begin{tabular}{cccccccccccccc}
\#&$n_s$ [Mpc$^{-3}$]& structure & observer & $B_{\rm obs}/$G &
$\alpha$ &${\cal L}^{40}_{l\leq10}$&${\cal L}^{40}_{l=1}$&
${\cal L}^{40}_{l\leq10}$&
${\cal L}^{40}_{l=1}$&${\cal L}^{10}_{l\leq10}$&${\cal L}^{10}_{l=1}$&
${\cal L}^{40}_{\theta\leq20^\circ}$&${\cal L}^{40}_{n\leq10}$\\
& & & & & & \multicolumn{2}{c}{AGASA only} &
\multicolumn{4}{c}{AGASA+SUGAR} & \multicolumn{2}{c}{AGASA only} \\
\hline \\
1 &$2.4\times10^{-4}$& yes & 1 & $1.3\times10^{-7}$ & 2.4 & 0.070 & 0.011
& 0.37 & 0.094 & 0.12 & 0.042 & 0.57 & 0.85 \\
2 &$2.4\times10^{-4}$& yes & 2 & $8.2\times10^{-12}$ & 2.4 & 0.43 & 0.35
& 0.52 & 0.48 & 0.16 & 0.18 & 0.52 & 0.85\\
3 &$2.4\times10^{-4}$& yes & 1 & 0 & 2.6 & 0.23 & 0.15 & 0.37
& 0.39 & 0.15 & 0.15 & 0.42 & 0.73 \\
4 &$2.4\times10^{-5}$& yes & 1 & 0 & 2.6 & 0.25 & 0.21 & 0.33
& 0.48 & 0.11 & 0.19 & 0.30 & 0.65 \\
5 &$2.4\times10^{-5}$& no & 1 & 0 & 2.6 & 0.36 & 0.34 & 0.45
& 0.51 & 0.13 & 0.24 & 0.65 & 0.71 \\
6 &$2.4\times10^{-5}$& yes & 2 & $8.2\times10^{-12}$ & 2.4 & 0.49 & 0.32
& 0.79 & 0.62 & 0.17 & 0.24 & 0.56 & 0.83 \\
7 &$2.4\times10^{-4}$& no & 1 & 0 & 2.6 & 0.35 & 0.40 & 0.42
& 0.47 & 0.12 & 0.17 & 0.53 & 0.78 \\
8 &$2.4\times10^{-6}$& no & 1 & 0 & 3.0 & 0.36 & 0.45 & 0.19
& 0.51 & 0.10 & 0.17 & 0.24 & 0.48 \\
9 &$2.4\times10^{-4}$& yes & 2 & 0 & 2.6 & 0.32 & 0.31 & 0.51
& 0.49 & 0.13 & 0.20 & 0.50 & 0.75 \\
10 &$2.4\times10^{-5}$& yes & 2 & 0 & 2.6 & 0.24 & 0.70 & 0.32
& 0.46 & 0.10 & 0.18 & 0.46 & 0.65 \\
\end{tabular}
\end{ruledtabular}
\end{table*}

Columns 7-14 of Tab.~\ref{tab1} show the likelihood significances 
discussed in the previous section obtained by comparing the predictions
from each scenario with currently available experimental data.
Particularly, as in previous studies,  we carry out a comparison in terms of  
multi-poles, auto-correlations and multi-plet statistics. 
As for the experimental data we use the 57 AGASA+Akeno events
above $4\times10^{19}\,$eV, when studying
the auto-correlation function and clustering properties of UHECRs
which are sensitive to small scales. 
For the large scale multi-poles, however, larger
sky coverage including the Southern hemisphere is desirable in order
to get realistic estimates of the true multi-poles.
Therefore, following Ref.~\cite{anchor_iso},
when comparing with multi-poles $l\leq10$
we will also use the combination of 
50 events observed by AGASA (excluding 7 events observed by Akeno)
and 49 events seen by SUGAR for a total of 99 events above 
$4\times10^{19}\,$eV.
In fact, as pointed out by the authors of Ref.~\cite{anchor_iso},
the AGASA and SUGAR experiments have comparable exposure in the
northern and southern hemisphere, respectively. 
In addition, while SUGAR's angular resolution is
much worse than for AGASA and in general prevents a combination of
the two data sets, this does not affect 
multi-poles $l\leq10$ because they are not sensitive to scales 
$\lesssim10^\circ$. Comparing columns 7 with 10 and 8 with 11 in
Tab.~\ref{tab1} shows that the inclusion of SUGAR data does not
significantly change the likelihood ranking of simulated scenarios
in terms of multi-poles.

Finally, using the combined exposure of AGASA+SUGAR, 
for each scenario we compute the expected values of the 
multi-pole coefficient for the $\simeq1500$ events observed above 
$10^{19}\,$eV. This is interesting because 
no sign of anisotropy was found by either the AGASA or the SUGAR
experiment at these energies.
Since data down to $10^{19}\,$eV are not publicly
available, we simply compare our results 
with a completely isotropic distribution.
The corresponding likelihoods are also summarized in Tab.~\ref{tab1}.

\begin{figure}
\includegraphics[width=0.49\textwidth,clip=true]{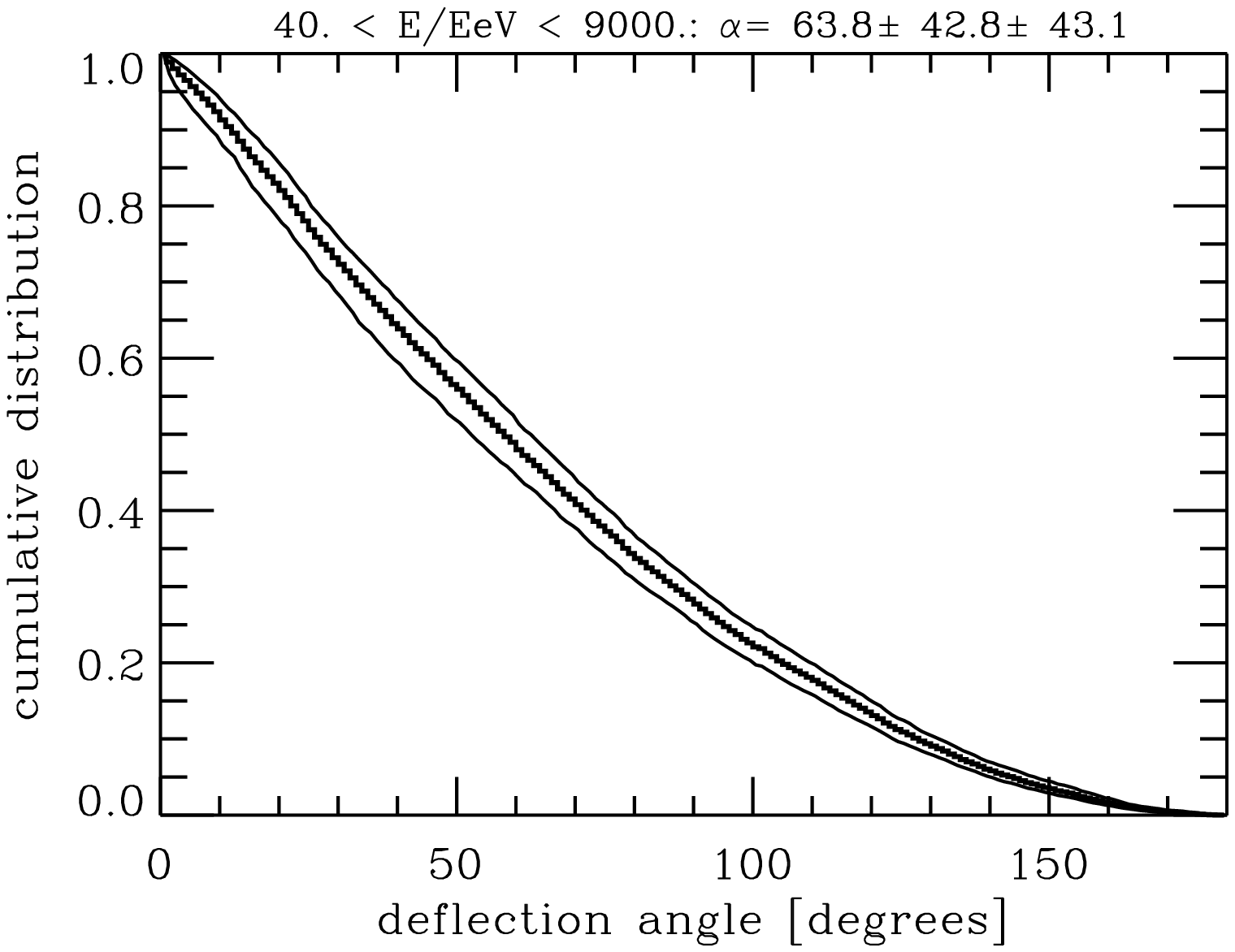}
\includegraphics[width=0.49\textwidth,clip=true]{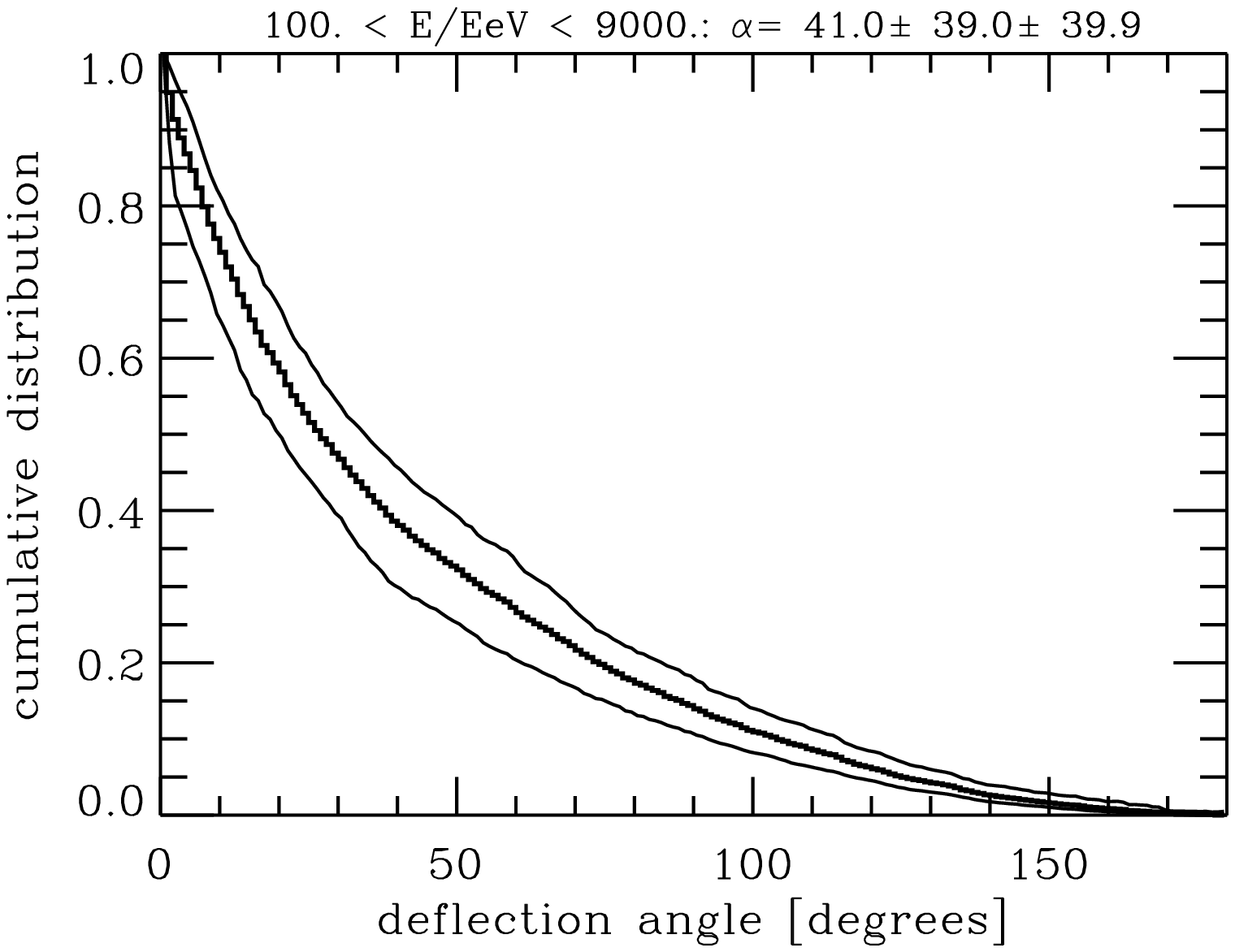}
\includegraphics[width=0.49\textwidth,clip=true]{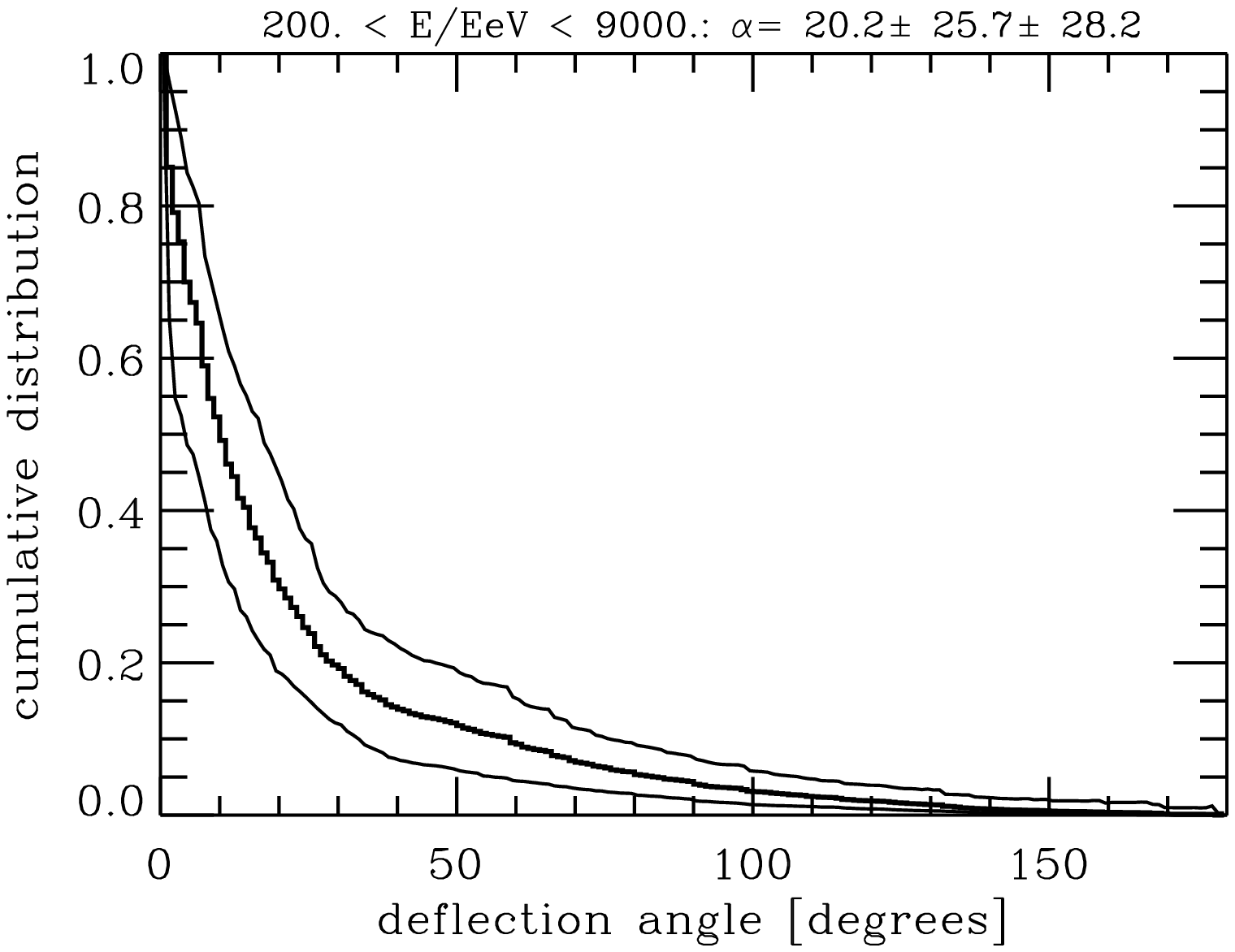}
\caption[...]{The cumulative distribution of UHECR deflection angles
with respect to the line of sight to the sources larger than $\alpha$ ,
averaged over 24 realizations for scenario 6 in Tab.~\ref{tab1}.
Shown are the distributions (middle, histogram) and 1-$\sigma$ variations
(upper and lower curves) above $4\times10^{19}\,$eV (upper panel), above
$10^{20}\,$eV (middle panel), and above $2\times10^{20}\,$eV
(lower panel). Also given on top of the figures average and
variances of the distributions.}
\label{fig5}
\end{figure}

As discussed in Ref.~\cite{letter}, the current best fit is provided by
scenario 6, i.e. for structured sources
of density $n_s\sim10^{-5}\,{\rm Mpc}^{-3}$ roughly following the
baryon density and immersed in fields up to a few micro Gauss,
whereas the observer is surrounded by fields $\ll10^{-7}\,$G.
In addition, we note that {\it if} the sources are homogeneously distributed,
the best fit density of the three values considered in Tab.~\ref{tab1}
(scenarios 5, 7, and 8) turns out to be the intermediate one,
$n_s\sim2.4\times10^{-5}$, in rough agreement with Ref.~\cite{bdm}.
However, the likelihood is a very shallow function of $n_s$ and
implies uncertainties of at least an order of magnitude.

To illustrate the general impact of an EGMF on propagation, we show in
Fig.~\ref{fig4} the distributions of delay times $\tau$ with respect
to straight line propagation with arrival energies $E$, averaged over all
realizations for scenario 6 in Tab.~\ref{tab1}. Note that the various
peaks at small time delays $\tau\lesssim10^{4}\,$yr are due to UHECRs
from discrete, nearby sources which mostly see the relatively weak
fields around Earth in this scenario. One can show that the number of
such peaks increases with the source density. For the same scenario 6,
Fig.~\ref{fig5} shows the distribution of UHECR deflection
angles $\alpha$ with respect to the line of sight to the sources above
various energy thresholds. This shows that deflection can be substantial
even at the highest energies. Qualitatively this can be understood
as follows: Neglecting energy loss processes, the r.m.s. deflection angle
over a distance $r$ in EGMF of r.m.s. strength $B$ and coherence length
$l_c$ is $\theta(E,r)\simeq(2rl_c/9)^{1/2}/r_L$~\cite{wm}, where the
Larmor radius of a particle of charge $Ze$ and energy $E$ is $r_L\simeq
E/(ZeB)$. In numbers this reads
\begin{eqnarray}
  \theta(E,r)&\simeq&0.8^\circ\,
  Z\left(\frac{E}{10^{20}\,{\rm eV}}\right)^{-1}
  \left(\frac{r}{10\,{\rm Mpc}}\right)^{1/2}\nonumber\\
  &&\hskip1cm\times\left(\frac{l_c}{1\,{\rm Mpc}}\right)^{1/2}
  \left(\frac{B}{10^{-9}\,{\rm G}}\right)\,,\label{deflec}
\end{eqnarray}
for $r\gtrsim l_c$. Keeping in mind that sources are correlated
with relatively strong fields and that according to Fig.~\ref{fig1}
$\simeq10$\% of the volume is filled with fields $\gtrsim10\,$nG,
we can see that deflections of order 20 degrees up to $10^{20}\,$eV
should therefore not be surprising. It is interesting to note in this
context that, as can be seen from Figs.~\ref{fig1}-\ref{fig3}, the
EGMF in our simulations tend to be more extended than the baryons
and thus the distribution of sources if they follow the baryons. A
significant part of the total deflection can therefore be contributed
by fields that are not in the immediate environment of the sources.
Furthermore, for the source density
$n_s\simeq2.4\times10^{-5}\,{\rm Mpc}^{-3}$ of the typical scenario
6 of Tab.~\ref{tab1}, the average distance to the closest source is
$\simeq15\,$Mpc, and thus close enough to make magnetized regions
of several Mpc around the sources extend of order ten degrees on the
sky, see also Fig.~\ref{fig3}, upper panel.

\begin{figure}[ht]
\includegraphics[width=0.49\textwidth,clip=true]{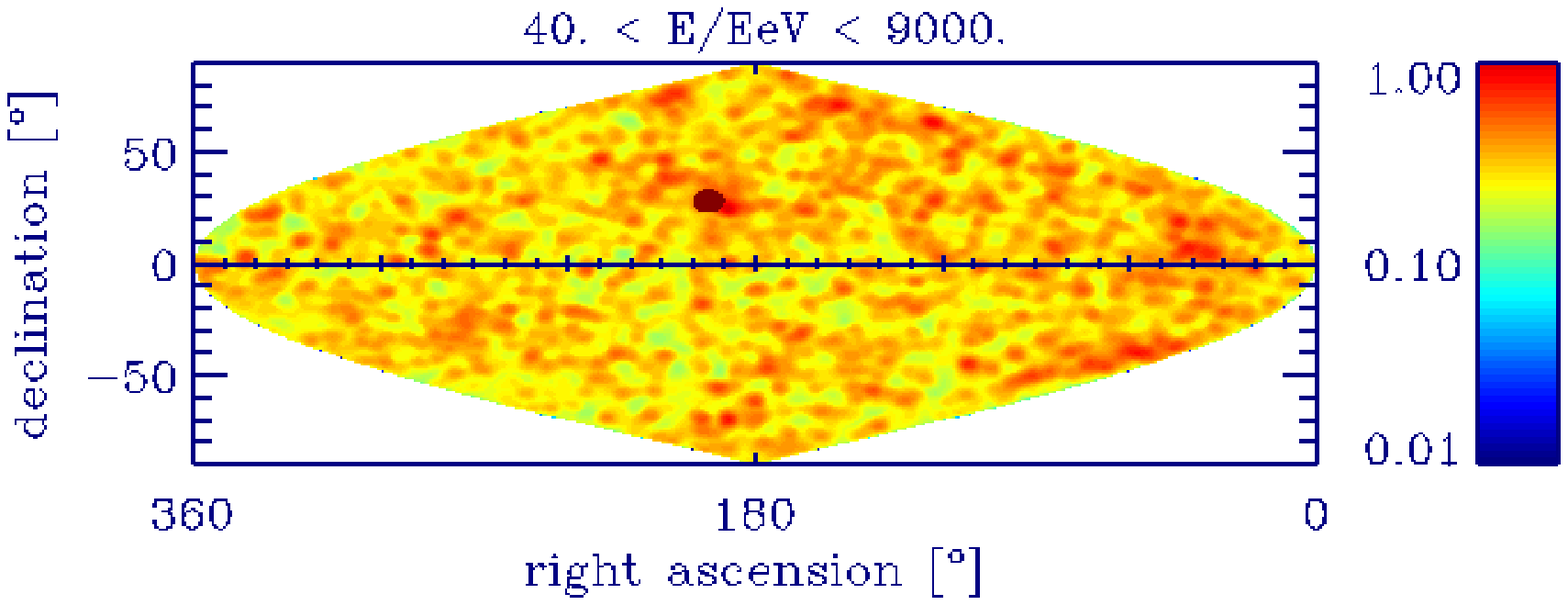}
\includegraphics[width=0.49\textwidth,clip=true]{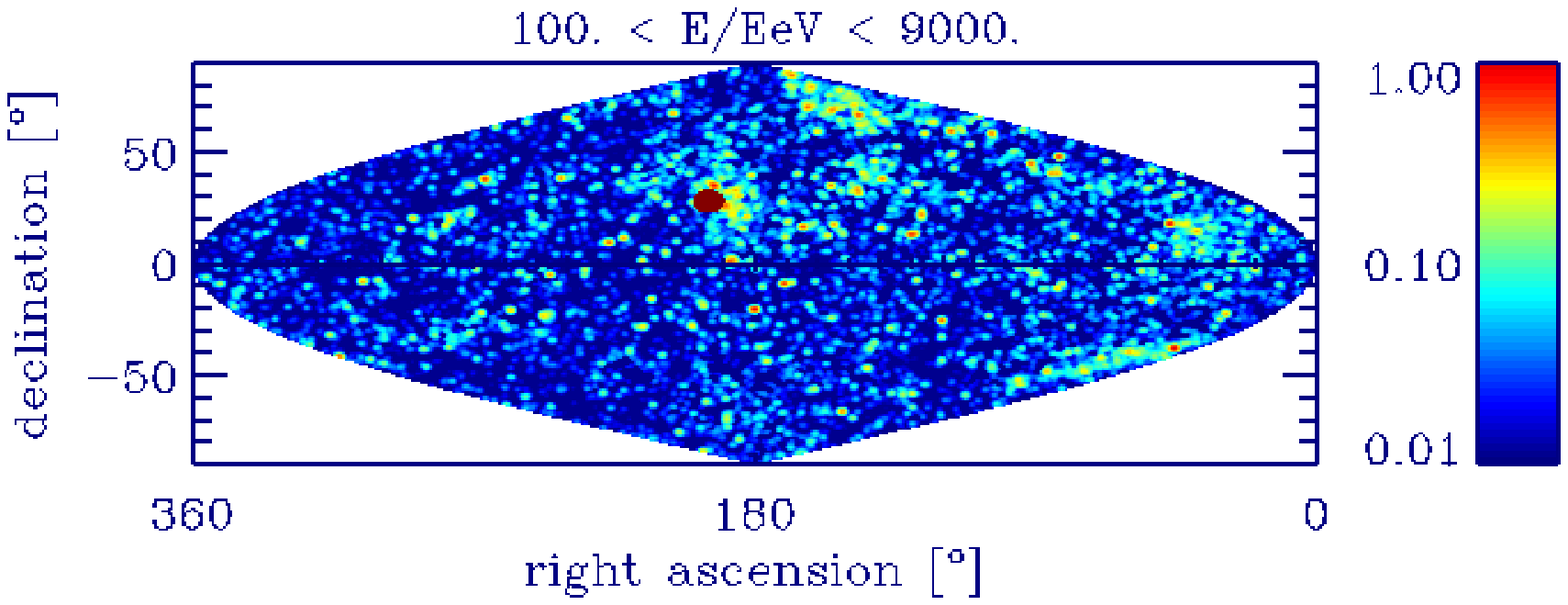}
\caption[...]{Arrival direction distributions predicted within
scenario 6 of Tab.~\ref{tab1}, averaged over 34 realizations
of $10^4$ simulated trajectories above $10^{19}\,$eV each.
The filled red sphere represents the overall direction to the
supergalactic centre. Upper panel: $\simeq10^5$ events above
$4\times10^{19}\,$eV, convoluted with an angular resolution of
$2.5^\circ$. Lower panel: $\simeq10^4$ events above
$10^{20}\,$eV, convoluted with an angular resolution of $1^\circ$.}
\label{fig6}
\end{figure}

The analytical approximation for the time delay corresponding to
the deflection Eq.~(\ref{deflec}) is $\tau(E,r)\simeq r\theta(E,r)^2/4$, or
\begin{eqnarray}
\tau(E,r)&\simeq&1.5 \times10^3\,Z^2
\left(\frac{E}{10^{20}\,{\rm eV}} \right)^{-2}
\left(\frac{r}{10\,{\rm Mpc}}     \right)^{2}\nonumber\\
&&\hskip1cm\times\left(\frac{l_c}{\rm Mpc}\right)
\left(\frac{B}{10^{-9}\,{\rm G}}  \right)^2 \, {\rm yr}\,.\label{tau}
\end{eqnarray}
These numbers are consistent with Fig.~\ref{fig4} if one takes into
account that the typical propagation distance drops from a few
hundred Mpc at $\sim10^{19}\,$eV to $\lesssim30\,$Mpc above
$\simeq10^{20}\,$eV. Finally, Fig.~\ref{fig6} shows,
for this same scenario, the predicted structure of arrival direction
distributions.

Sufficient data to definitely discriminate among the different
scenarios presented in Tab.~\ref{tab1}, e.g. the presence of diffuse EGMF
as well as UHECR source characteristics,
will have to await the next-generation experimental facilities, 
such as the Pierre Auger~\cite{auger} and EUSO~\cite{euso}
projects. To demonstrate what can be achieved with these new experiments,
we will work out the predictions for various observable statistical quantities 
for UHECRs both above $4\times10^{19}\,$eV
and $10^{20}\,$eV for the Auger full sky observatory. For this we will
assume an exposure about 25 times larger than AGASA, which is reachable 
after about three years of running experiment. It corresponds to
$\sim1500$ events observed above $4\times10^{19}\,$eV and
$\sim30$ events above $10^{20}\,$eV if a GZK cutoff is present,
or $\sim200$ events above $10^{20}\,$eV if no GZK cutoff is present.
We will also sometimes consider $\sim5000$ events above
$4\times10^{19}\,$eV, as is expected to be easily achieved by
the EUSO experiment~\cite{euso}.

\subsection{Signatures of structured versus homogeneous source distributions}

\begin{figure}[ht]
\includegraphics[width=0.49\textwidth,clip=true]{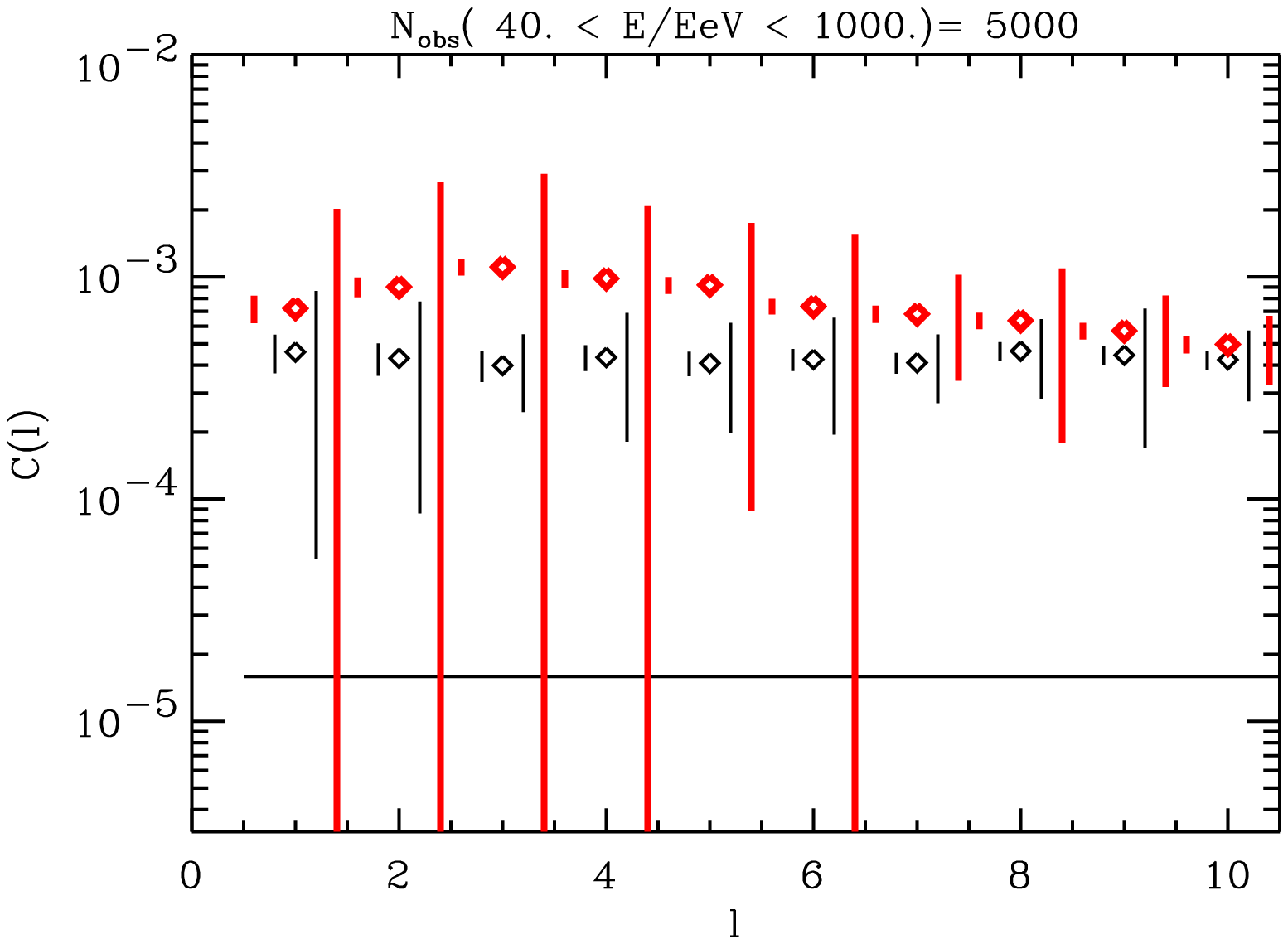}
\includegraphics[width=0.49\textwidth,clip=true]{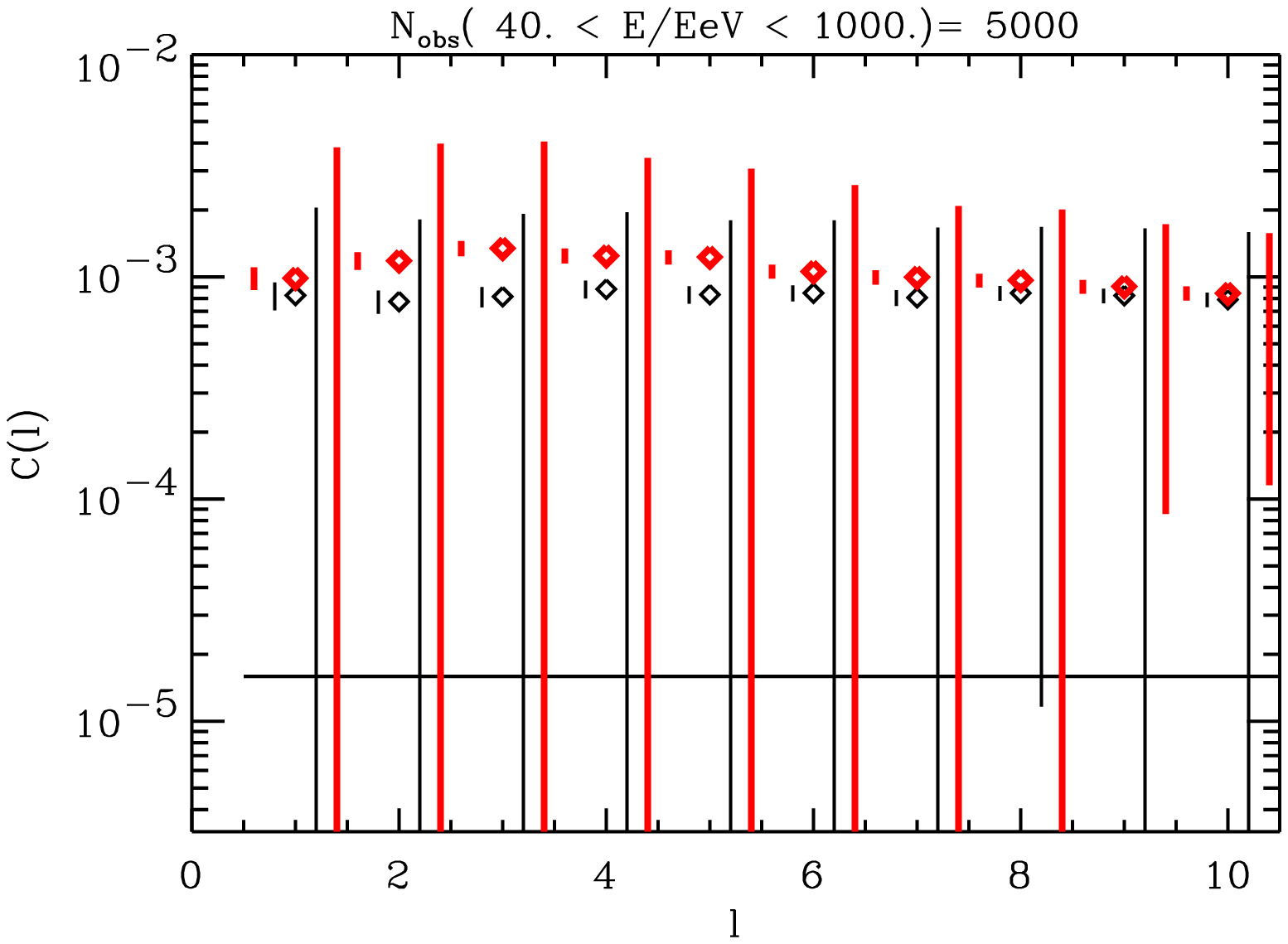}
\caption[...]{Angular power spectrum $C(l)$ as a function of
multi-pole $l$, for the full sky Pierre Auger observatory
assuming $N_{\rm obs}=5000$ events observed above $4\times10^{19}\,$eV.
The upper panel assumes sources with identical properties,
whereas the luminosity function Eq.~(\ref{source_prop}) was used
in the lower panel. We show the realization averages (diamonds),
statistical (left) and total (right) error bars, respectively,
predicted by the model. The red (thick) diamonds and red (thick, outer)
error bars represent scenario 4 (unmagnetized structured sources),
whereas the black (thin) diamonds and black (thin, inner) error
bars represent scenario 5 (unmagnetized homogeneously distributed
sources). The source
density is $n_s=2.4\times10^{-5}\,{\rm Mpc}^{-3}$ in both cases.
The straight line is the analytical prediction,
$C_l\simeq(4\pi N_{\rm obs})^{-1}$, for the average multi-poles for
complete isotropy.}
\label{fig7}
\end{figure}

\begin{figure}[ht]
\includegraphics[width=0.49\textwidth,clip=true]{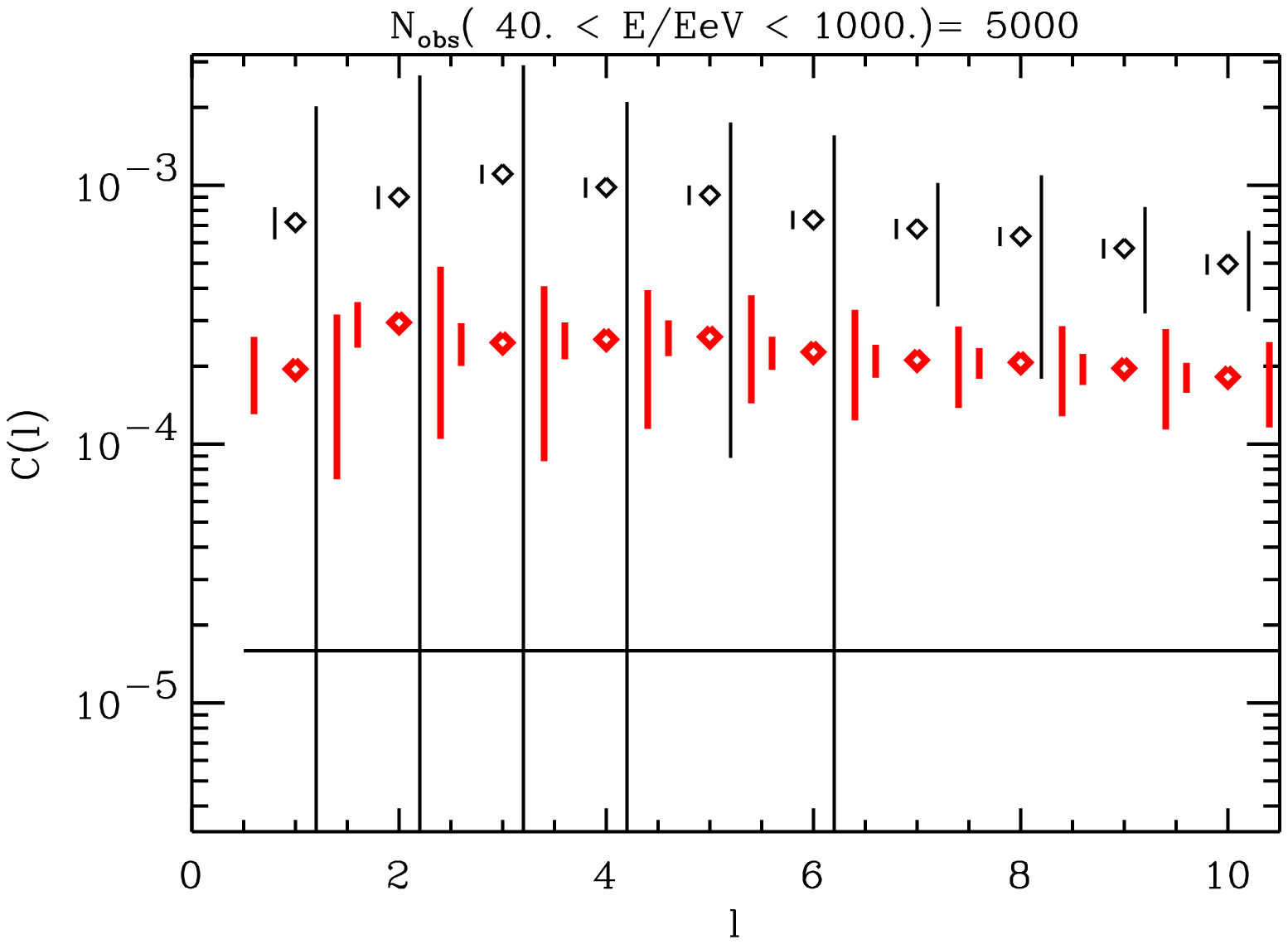}
\includegraphics[width=0.49\textwidth,clip=true]{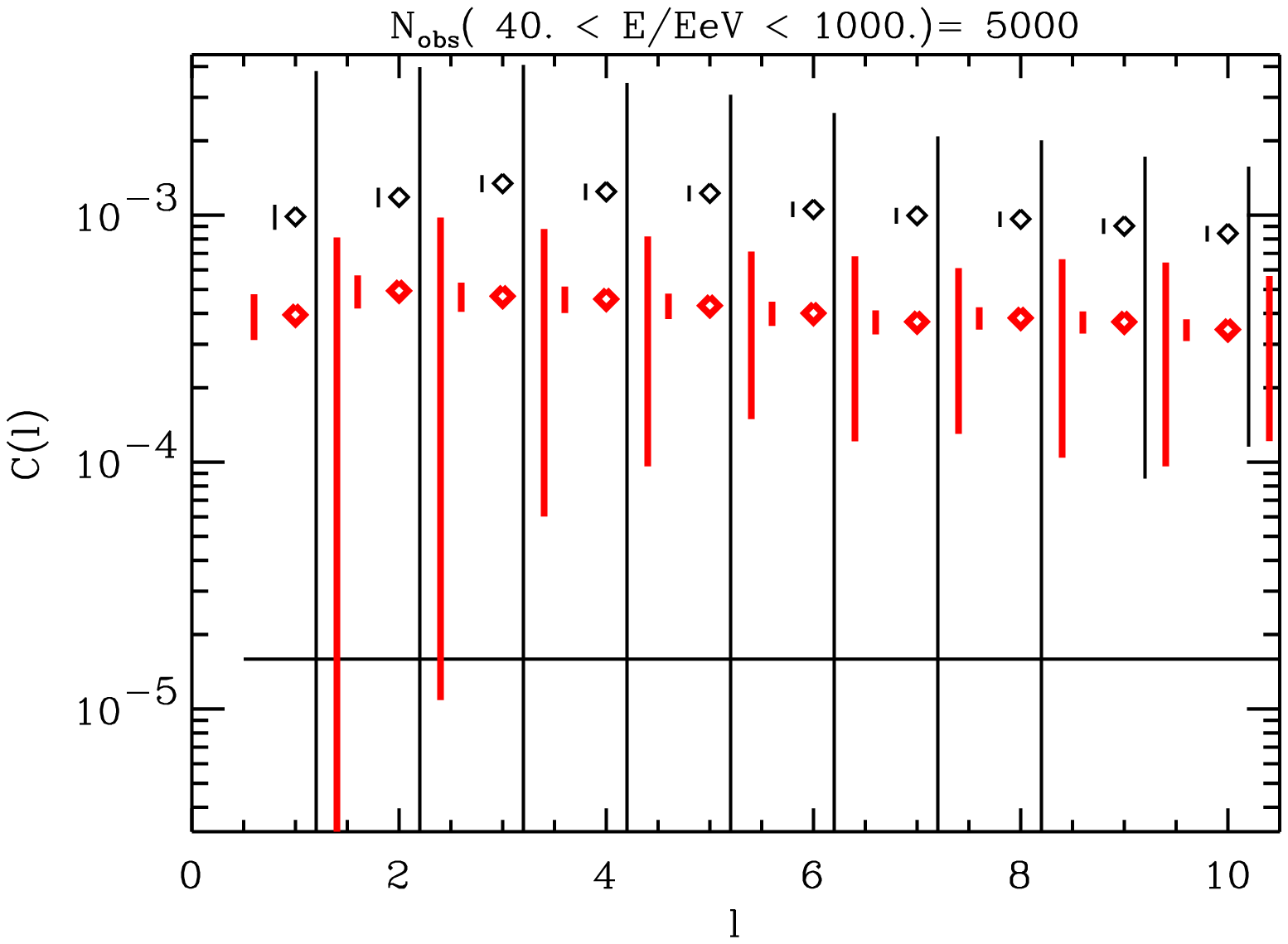}
\caption[...]{Same as Fig.~\ref{fig7}, but comparing predictions
of structured sources with same densities and different magnetizations.
The red (thick) diamonds and red (thick, outer) error bars represent
scenario 6 (magnetized sources),
whereas the black (thin) diamonds and black (thin, inner) error bars
represent scenario 4 (unmagnetized sources).}
\label{fig8}
\end{figure}

One expects that the nonuniformity in the source distribution
({\it structure}) mostly influences the large scale multi-poles.
Fig.~\ref{fig7} shows that, for a given source density,
if magnetic deflection can be neglected, 
a homogeneous distribution of sources predicts, not surprisingly,
an angular power
spectrum more isotropic than a structured source distribution.
Unfortunately, cosmic variance is sufficiently
large that one can definitely discriminate the case of a structured
distribution of sources only if the measured multi-poles happen to be several
standard deviations above the prediction of a homogeneous distribution.
This can happen only when the source luminosities are almost uniform,
which is rather unrealistic. Otherwise, 
the measured multi-poles are consistent
with both structured and unstructured source distributions, see
Fig.~\ref{fig7}. In addition, within cosmic variance, the
multi-poles depend insignificantly on the source density.
In contrast, magnetic fields tend to significantly decrease
their values, while at the same time reducing cosmic variance,
see Fig.~\ref{fig8}. As a result, predictions of low-scale multi-poles
above $4\times10^{19}\,$eV for next generation experiments tend to
deviate from isotropy more significantly than for unmagnetized sources.

\begin{table}[ht]
\caption[...]{Predictions for the sum of the first ten multi-poles,
$\sum_{l=1}^{10}C_l$, above $4\times10^{19}\,$eV for $N_{\rm obs}=5000$
observed events, with and without source luminosity variations
according to Eq.~(\ref{source_prop}). The two errors represent
statistical and total (including cosmic variance) errors, as in
the figures. The rows represent the scenarios of Tab.~\ref{tab1}.}
\label{tab2}
\begin{ruledtabular}
\begin{tabular}{ccc}
 & luminosity function Eq.~(\ref{source_prop}), & constant luminosity, \\
 & $10^3\sum_{l=1}^{10}C_l$ & $10^3\sum_{l=1}^{10}C_l$ \\
\hline
2 & 6.87$\pm$0.39$\pm$5.9 & 2.21$\pm$0.16$\pm$0.58 \\
3 & 10.6$\pm$0.52$\pm$17 & 4.48$\pm$0.27$\pm$2.3 \\
4 & 10.8$\pm$0.52$\pm$16 & 7.75$\pm$0.43$\pm$8.5 \\
5 & 8.24$\pm$0.46$\pm$9.0 & 4.30$\pm$0.28$\pm$1.7 \\
6 & 4.11$\pm$0.25$\pm$2.7 & 2.27$\pm$0.17$\pm$0.69 \\
7 & 8.30$\pm$0.46$\pm$9.4 & 2.57$\pm$0.18$\pm$0.55 \\
8 & 14.8$\pm$0.75$\pm$16 & 14.9$\pm$0.73$\pm$16 \\
9 & 9.07$\pm$0.49$\pm$13 & 2.88$\pm$0.20$\pm$0.57 \\
10 & 11.4$\pm$0.58$\pm$12 & 8.38$\pm$0.48$\pm$5.1 \\
\end{tabular}
\end{ruledtabular}
\end{table}

To demonstrate these tendencies more quantitatively, in Tab.~\ref{tab2}
we provide the predictions for the sum of the first ten multi-poles
for the scenarios from Tab.~\ref{tab1}.

\subsection{Source Density}
\begin{figure}[ht]
\includegraphics[width=0.49\textwidth,clip=true]{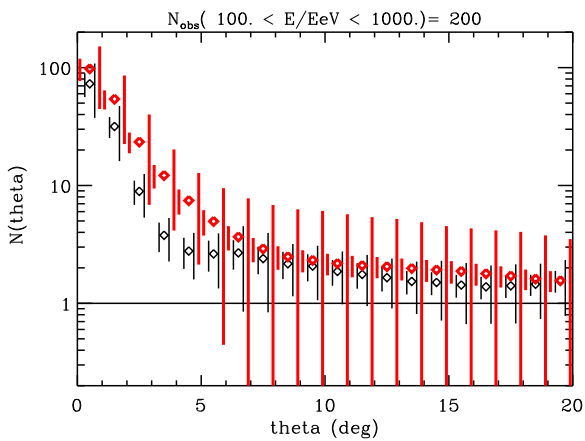}
\caption[...]{The auto-correlation function for $N_{\rm obs}=200$ events
observed above $10^{20}\,$eV for the full sky Pierre Auger observatory
for a bin size of $\Delta\theta=1^\circ$. Averages and errors are as
in Fig.~\ref{fig7}. Compared are unmagnetized structured sources with
different densities: The red (thick) diamonds and red (thick, outer)
error bars represent scenario 4 ($n_s=2.4\times10^{-5}\,{\rm Mpc}^{-3}$),
whereas the black (thin) diamonds and black (thin, inner) error bars
represent scenario 3 ($n_s=2.4\times10^{-4}\,{\rm Mpc}^{-3}$) in
Tab.~\ref{tab1}.}
\label{fig9}
\end{figure}

\begin{figure}[ht]
\includegraphics[width=0.49\textwidth,clip=true]{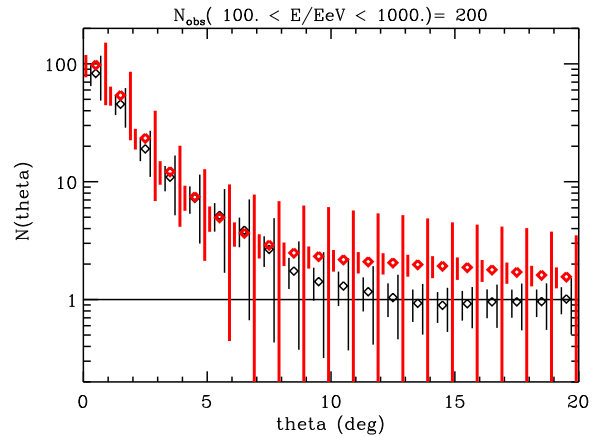}
\caption[...]{Same as Fig.~\ref{fig9}, but comparing unmagnetized
sources with the same density $n_s=2.4\times10^{-5}\,{\rm Mpc}^{-3}$
with structured and unstructured distributions: The red (thick)
diamonds and red (thick, outer) error bars represent scenario 4
(structured source distribution), whereas the black (thin) diamonds
and black (thin, inner) error bars represent scenario 5 (homogeneous
source distribution) in Tab.~\ref{tab1}.}
\label{fig10}
\end{figure}

If magnetic field effects can be neglected, the small-scale
auto-correlation function will depend strongly on the source
density: Few sources imply strong auto-correlations with considerable
cosmic variance as well as strong clustering, whereas many sources
imply weak auto-correlation with comparatively small cosmic
variance. In the case of homogeneously distributed sources with
identical properties
this was indeed suggested as a measure of the source density~\cite{bdm}.
Fig.~\ref{fig9} demonstrates that this is also true for structured
sources. Furthermore, for a given source density the small scale
auto-correlation is relatively insensitive to the structure of the
source distribution, as becomes evident in Fig.~\ref{fig10}.

However, if the sources are immersed in magnetic fields
of order $\gtrsim0.1\mu$G, as they can occur in galaxy clusters
and filaments, the auto-correlation function becomes almost
insensitive to the source density and instead becomes a probe
of source magnetization, as discussed below in Sect.~3.D. This is
because a structured EGMF of such strength diffuses cosmic rays
up to $10^{20}\,$eV over the whole region immersed in such fields,
as can be seen from the rough estimate Eq.~(\ref{deflec}). Therefore,
the number of sources within such regions doesn't significantly affect
the auto-correlations any more, which increases the uncertainty in the
source density.

\subsection{Source Luminosity Function}

\begin{figure}[ht]
\includegraphics[width=0.49\textwidth,clip=true]{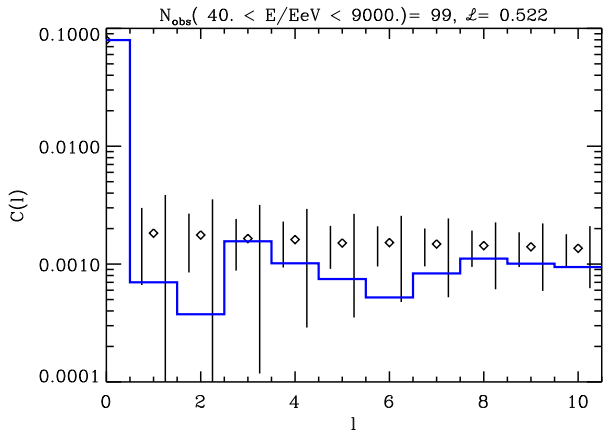}
\includegraphics[width=0.49\textwidth,clip=true]{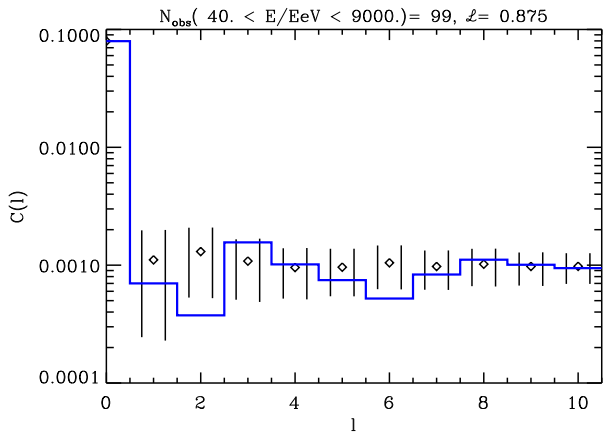}
\caption[...]{The angular power spectrum $C(l)$ as a function of
multi-pole $l$, predicted for the AGASA+SUGAR exposure function (see text),
for $N_{\rm obs}=99$ events observed above 40 EeV, sampled from the simulated
configurations of scenario 2 in Tab.~\ref{tab1}. The diamonds indicate
the realization averages, and the left and right error
bars represent the statistical and total (including cosmic variance
due to different realizations)
error, respectively, see text for explanations. The histogram
represents the AGASA+SUGAR data. The overall likelihood significance
for $n=4$ and $l\leq10$ in Eq.~(\ref{chi_n}) appears at the top of the
figures. The upper panel takes into account cosmic variance due to variation
of source properties assumed to be parameterized by Eq.~(\ref{source_prop}).
The lower panel assumes that all sources have identical properties
and thus cosmic variance is uniquely due to variation in source location.}
\label{fig11}
\end{figure}

If the source luminosity function is parametrized by
$dn_s/dQ\propto Q^{-\alpha}$ in an interval 
$Q_{\rm min}\leq Q\leq Q_{\rm max}$,
then for $\alpha\ll2$ the most luminous sources will dominate,
whereas for $\alpha\gg2$ the weakest sources will dominate. These
two cases will therefore be approximated by scenarios with sources
of identical properties and cosmic variance is expected to be
minimal. In contrast, $\alpha\simeq2$ implies that all luminosities
contribute approximately equally to the observed flux. This
is close to the case we chose in Eq.~(\ref{source_prop}) which
therefore tends to maximize cosmic variance.
As can be seen from Fig.~\ref{fig11}, in this case
the uncertainties of large scale multi-poles due to
cosmic variance are in general larger than the ones due to the
finite number of events observed, even for the relatively sparse
data set currently available. For the small scale
properties described by the auto-correlation function and cluster
frequencies, for $N_{\rm obs}\lesssim100$, cosmic variance is in general
smaller than or comparable to the statistical error, as can be seen
from Figs.~\ref{fig9} and~\ref{fig10} and Figs.~\ref{fig12}
and~\ref{fig13} below.

\begin{figure}[ht]
\includegraphics[width=0.49\textwidth,clip=true]{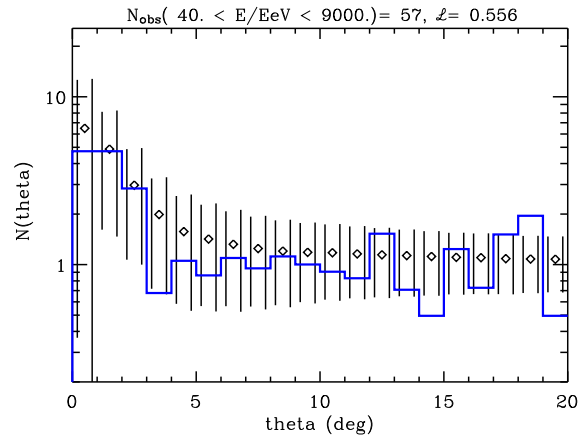}
\includegraphics[width=0.49\textwidth,clip=true]{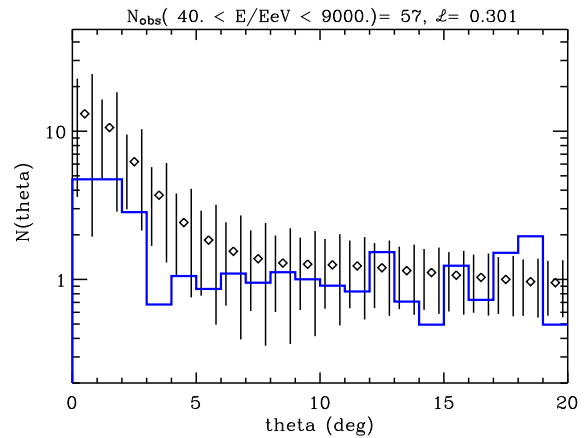}
\caption[...]{The angular correlation function $N(\theta)$ as a function
of angular distance $\theta$, predicted for a bin size of
$\Delta\theta=1^\circ$ for the AGASA exposure function (see text),
for $N_{\rm obs}=57$ events observed above 40 EeV, sampled from the simulated
configurations. Note that an isotropic distribution would correspond
to $N(\theta)=1$. Averages and errors are as in Fig.~\ref{fig7}.
The histogram represents the AGASA data.
The two scenarios shown correspond to structured sources of equal
density with (scenario 6, upper panel) and without (scenario 4, lower panel)
magnetization. The overall likelihood significance for $n=4$
and $\theta\leq10^\circ$ in Eq.~(\ref{chi_n}) appears at the top of the
figures. It is not significantly different for somewhat
larger bin sizes $\Delta\theta\simeq2^\circ$.}
\label{fig12}
\end{figure}

\subsection{Source Magnetization}
Scenarios 3 and 4 are somewhat disfavored by auto-correlation
and clustering: Structured sources produce more clustering in the
absence of magnetic fields, as can be seen, for example, in
Fig.~\ref{fig12}. While this effect is marginal in the current
data set, it will develop into a strong discriminator for future
experiments:
Scenarios with no significant magnetic fields predict a stronger
auto-correlation at small angles, independent of whether or not
the sources are structured. This is because if sources are immersed
in considerable magnetic fields, their images are smeared out,
which also smears out the auto-correlation function over a few
degrees.

\begin{figure}[ht]
\includegraphics[width=0.49\textwidth,clip=true]{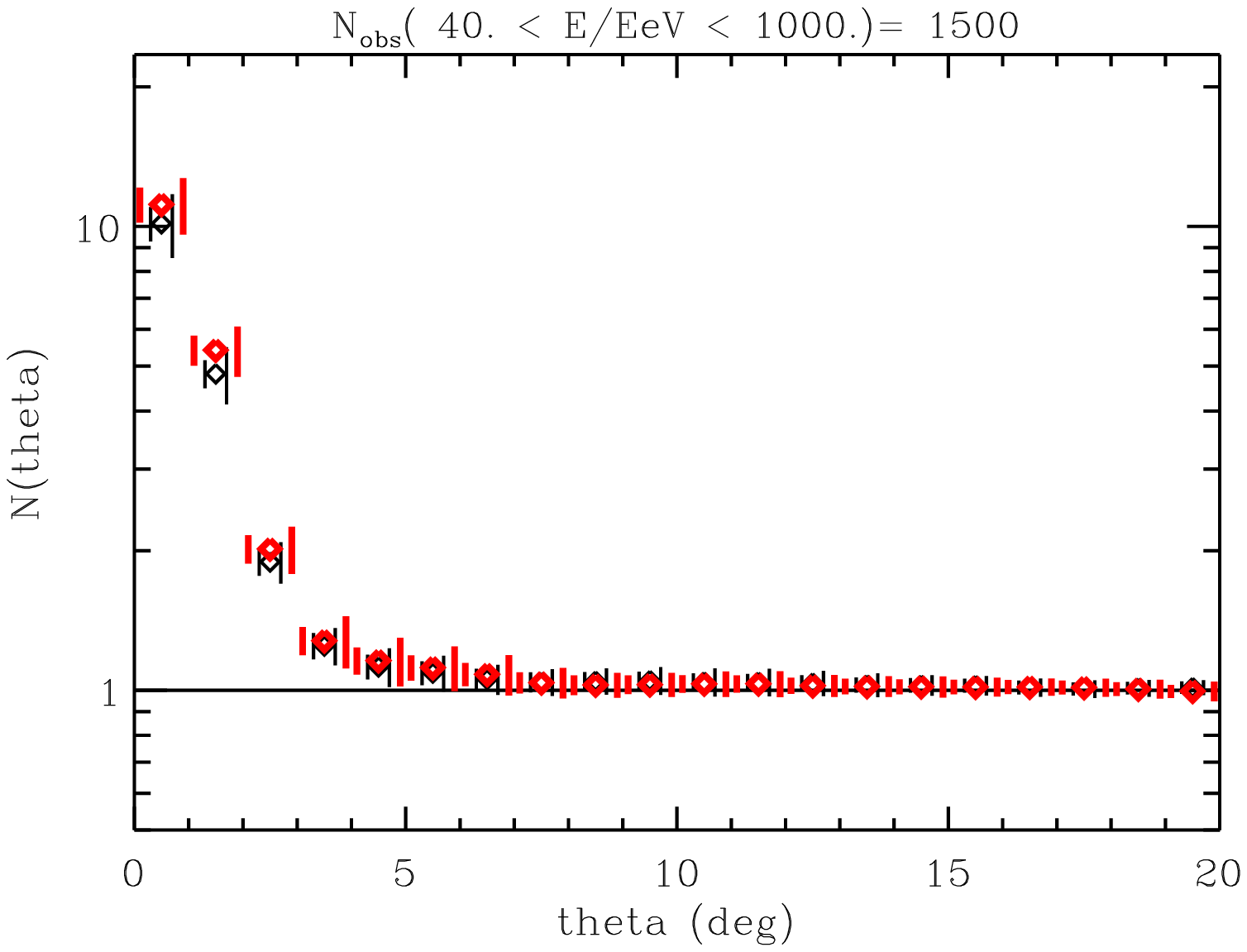}
\includegraphics[width=0.49\textwidth,clip=true]{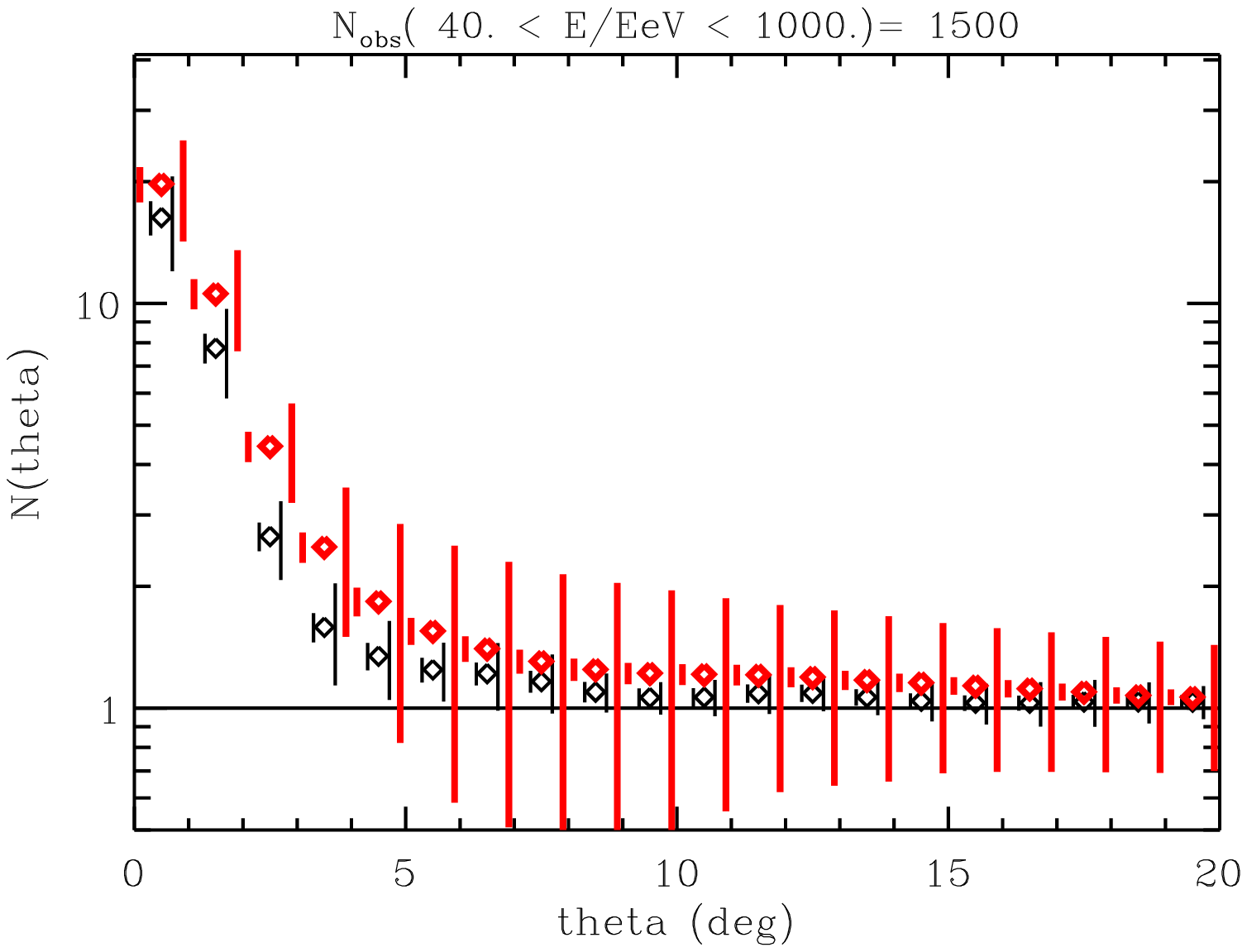}
\caption[...]{The auto-correlation function for $N_{\rm obs}=1500$ events
observed above 40 EeV for the full sky Pierre Auger observatory.
Upper panel: Comparing magnetized sources with
different densities. The red (thick) diamonds and red (thick, outer)
error bars represent scenario 6 ($n_s=2.4\times10^{-5}\,{\rm Mpc}^{-3}$),
whereas the black (thin) diamonds and black (thin, inner) error bars
represent scenario 2 ($n_s=2.4\times10^{-4}\,{\rm Mpc}^{-3}$) in\
Tab.~\ref{tab1}. Lower panel: Same as upper panel, but in the absence
of magnetic fields, i.e. comparing scenario 4 (red, thick set) with
scenario 3 (black, thin set) in Tab.~\ref{tab1}.}
\label{fig13}
\end{figure}

As shown in Fig.~\ref{fig13}, this effect is significant once
$N_{\rm obs}\gtrsim10^3$ events are observed above $4\times10^{19}\,$eV,
as will be the case for the Pierre Auger project, and also
if $N_{\rm obs}\gtrsim100$ events are observed above $10^{20}\,$eV, as could
be achieved, for example, by the EUSO project. At these energies
the strongest auto-correlation at small angles is predicted by
scenario 4 in Tab.~\ref{tab1}, where sources follow the large scale
structure in the absence of magnetic fields. For a given source density,
the auto-correlation predicted by homogeneously distributed sources
in the absence of magnetic fields
(scenario 5 in Tab.~\ref{tab1}) lies in between the one predicted
by structured sources in the absence of magnetic fields (scenario
4 in Tab.~\ref{tab1}) and the one of structured sources with magnetic
fields (scenario 6 in Tab.~\ref{tab1}).

\begin{table}[ht]
\caption[...]{Predictions for the auto-correlation function
in the first degree bin, $N(1^\circ)$, for various threshold
energies and number of observed events $N_{\rm obs}$. The two errors represent
statistical and total (including cosmic variance) errors, as in
the figures. The rows represent the scenarios of Tab.~\ref{tab1}.}
\label{tab3}
\begin{ruledtabular}
\begin{tabular}{ccc}
 & $E\geq10^{20}\,$eV, & $E\geq4\times10^{19}\,$eV, \\
\# & $N_{\rm obs}=200$ & $N_{\rm obs}=1500$ \\
\hline
2 & 45.4$\pm$12$\pm$17 & 10.1$\pm$0.86$\pm$1.6 \\
3 & 72.9$\pm$17$\pm$35 & 16.3$\pm$1.6$\pm$4.3 \\
4 & 98.0$\pm$21$\pm$53 & 19.7$\pm$2.0$\pm$5.5 \\
5 & 83.0$\pm$18$\pm$34 & 15.9$\pm$1.5$\pm$3.4 \\
6 & 44.8$\pm$12$\pm$19 & 11.2$\pm$0.97$\pm$1.6 \\
7 & 60.8$\pm$15$\pm$30 & 13.1$\pm$1.3$\pm$2.2 \\
8 & 191$\pm$33$\pm$88 & 37.9$\pm$3.1$\pm$11.6 \\
9 & 68.6$\pm$17$\pm$37 & 14.4$\pm$1.4$\pm$3.6 \\
10 & 100$\pm$21$\pm$40 & 20.2$\pm$1.9$\pm$4.4 \\
\end{tabular}
\end{ruledtabular}
\end{table}

\begin{figure}[ht]
\includegraphics[width=0.49\textwidth,clip=true]{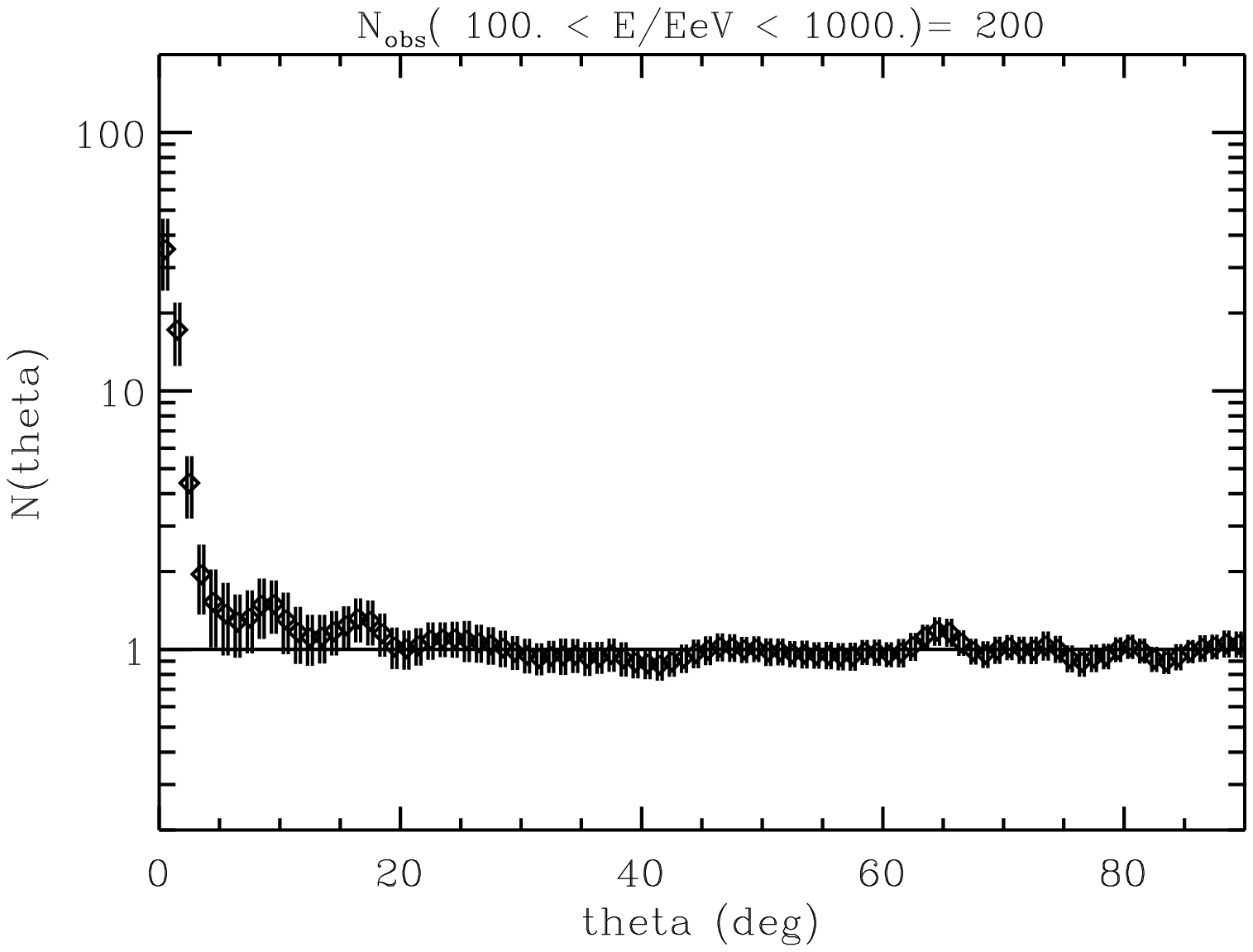}
\includegraphics[width=0.49\textwidth,clip=true]{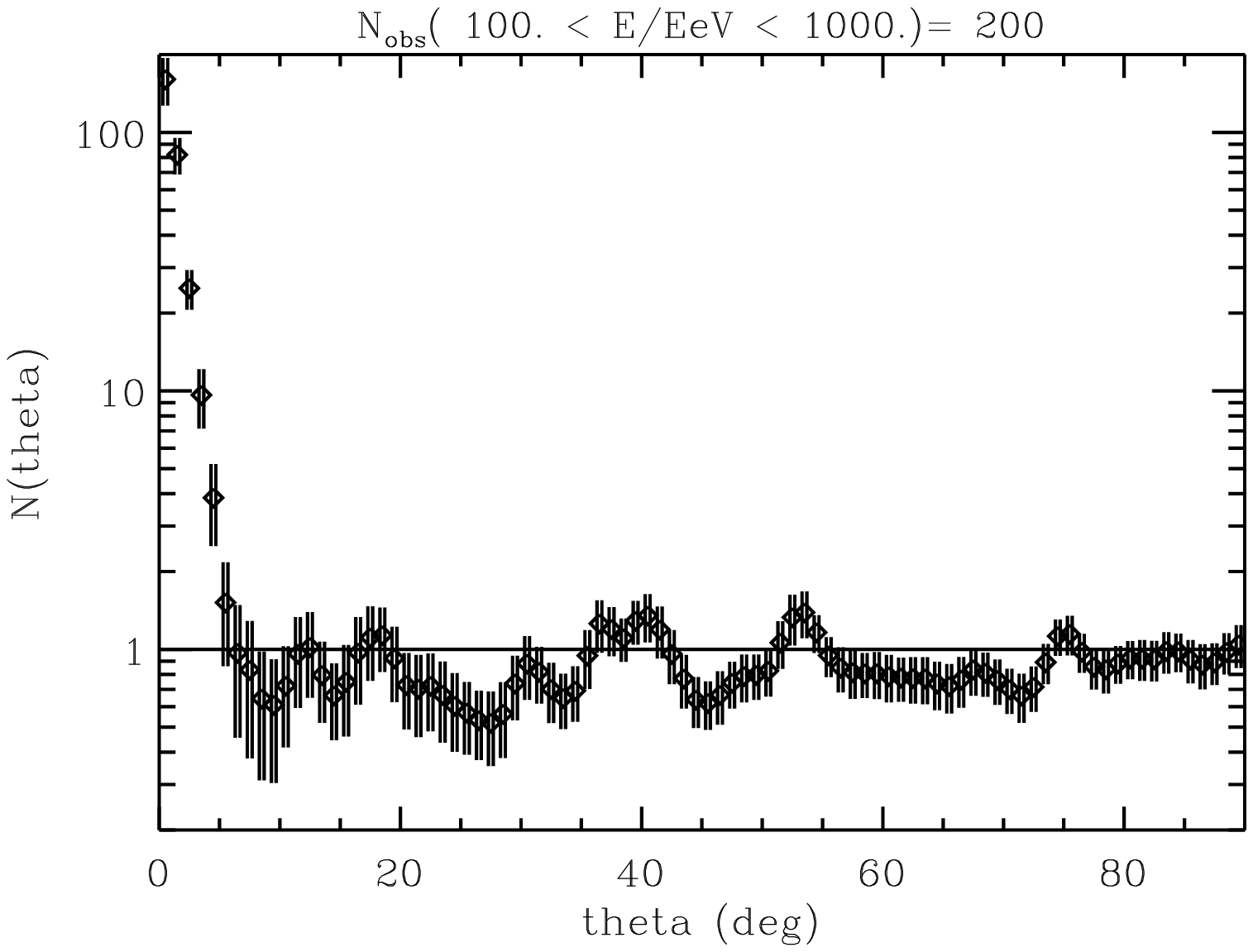}
\caption[...]{The auto-correlation function for the full sky Pierre
Auger observatory, for $N_{\rm obs}=200$ events
observed above 100 EeV, predicted by one particular source
realization for $n_s=2.4\times10^{-5}\,{\rm Mpc}^{-3}$.
Upper panel: scenario 6 in Tab.~\ref{tab1}, i.e. with magnetic fields.
Lower panel: scenario 4 in Tab.~\ref{tab1}, i.e. without magnetic fields.
\label{fig14}}
\end{figure}

As an example, the predictions for the auto-correlation function
in the first degree bin are summarized in Tab.~\ref{tab3}. This
also confirms the discussion in Sect.~3.B: For structured sources
in the absence of
magnetic fields, the degree-scale auto-correlation function
decreases with increasing source density, whereas it is rather
independent of the source density if the sources are immersed
in magnetic fields. Finally we observe that the low-$l$ auto-correlation
in case of magnetized sources hardly depends on their densities,
as can be seen by comparing scenarios 2 and 6 in Tab.~\ref{tab3}.
In contrast, for unmagnetized sources the degree-scale auto-correlation
is quite sensitive to the source density and always significantly
higher than for magnetized discrete sources. This is clearly
demonstrated by Fig.~\ref{fig13}. A consequence of this is that
in the absence of magnetic fields the auto-correlation function
and its fluctuations can indeed be used as a measure of the
source density, as suggested in Ref.~\cite{bdm}: The strength
of the auto-correlation at small scales as well as its cosmic
variance at all scales increase with decreasing source density.
However, this effect can be almost completely erased by magnetic
fields surrounding the sources. Such fields also tend to considerably
reduce the effects of cosmic variance, especially at low source densities,
as Fig.~\ref{fig13} demonstrates. This is also reflected by the
fact that oscillations predicted to occur in the actually measured
auto-correlation function, i.e. predicted for a given realization
of sources and magnetic fields, are considerably suppressed by
magnetic fields. This is demonstrated by Fig.~\ref{fig14}.

Furthermore, Tab.~\ref{tab3} shows that the small-scale
auto-correlation is rather independent of the observer position:
The pairs of simulations 3, 9 and 4, 10 with unmagnetized sources
differ only in the observer position, respectively, see Tab.~\ref{tab1}.
Their predictions for the average degree-scale auto-correlations
indeed differ by amounts much smaller than even the statistical
error. This suggests that the concrete realization of the local
large scale structure is mostly irrelevant to tendencies presented
in the current paper.

As discussed in Sect.~3A, magnetized sources also tend to suppress
large scale multi-poles as well as their cosmic variance. However,
since unmagnetized sources, in particular in case of structured
distributions, predict cosmic variances nearly as large as the
averages of the multi-poles, small multi-poles cannot be used
as a signature of strongly magnetized sources.

\subsection{Magnetic Fields surrounding the Observer}
\begin{figure}[ht]
\includegraphics[width=0.49\textwidth,clip=true]{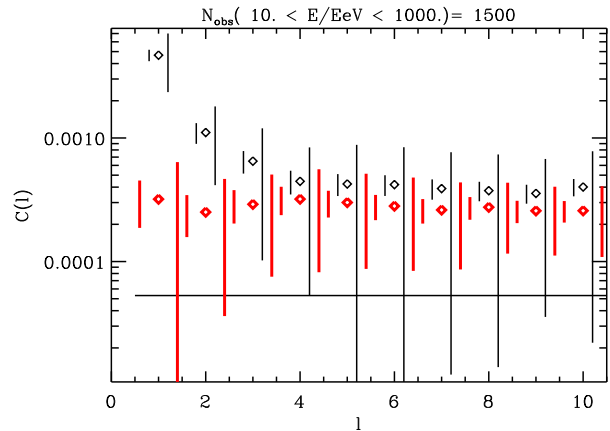}
\caption[...]{The angular power spectrum obtained for the combined
AGASA+SUGAR exposure function, for $N_{\rm obs}=1500$ events observed
above 10 EeV. The red (thick) diamonds and red (thick, outer)
error bars represent scenario 6, whereas the black (thin) diamonds
and black (thin, inner) error bars represent scenario 1. The line
key is as in Fig.~\ref{fig7}.}
\label{fig15}
\end{figure}

If the observer is immersed in $\sim0.1\mu\,$G fields,
considerable low-$l$ multi-poles are predicted. This possibility is already
disfavored by current data~\cite{letter}, as seen from Tab.~\ref{tab1}.
We demonstrate this by comparing the predictions of scenarios
with a strongly or negligibly magnetized observer for the
large scale power spectrum above $10^{19}\,$eV in Fig.~\ref{fig15}.
This can qualitatively be understood as follows~\cite{letter}:
The presence of a region with a relatively strong magnetic 
field surrounding the observer preferentially shields off UHECRs
from sources that are farther away from the observer, in particular
those outside of the magnetized region. In contrast, the flux from
sources within the magnetized region is enhanced by the increased
confinement time. Thus the observed flux
is dominated by a few closer sources and appears more anisotropic.

\begin{figure}[ht]
\includegraphics[width=0.49\textwidth,clip=true]{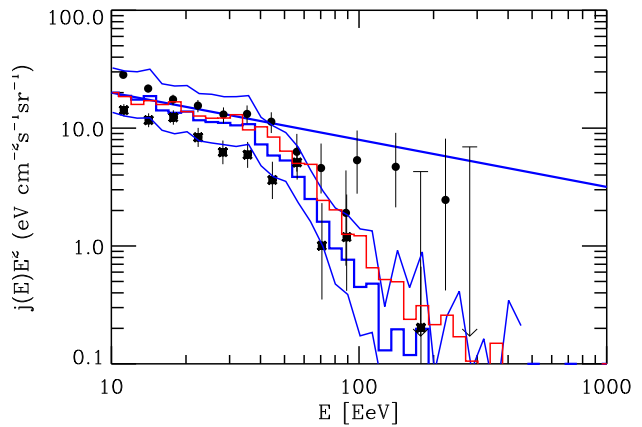}
\caption[...]{Predicted spectrum observable by AGASA for scenario
2 (weakly magnetized observer), as compared to the AGASA (dots) and
HiRes-I~\cite{hires} (stars) data. The histogram marks the
average and the two lines above and below the 1-sigma fluctuations
over the simulated realizations. The solid straight line marks the
injection spectrum. For comparison shown (red, thin histogram)
is the spectrum predicted by scenario 1 (strongly magnetized
observer). Both cases correspond to the same source density
$n_s=2.4\times10^{-4}\,{\rm Mpc}^{-3}$.}
\label{fig16}
\end{figure}

Furthermore, scenarios in which the observer is immersed in
an EGMF $\ll0.1\mu\,$G predict UHECR spectra with a pronounced
GZK cutoff. In contrast, if the observer is strongly magnetized
the cutoff is attenuated. One can understand this
as a partial compensation between energy attenuation lengths
which decrease with increasing energy and magnetic diffusion
lengths which increase with energy. As long as the source
density is sufficiently high around the observer, the number
of sources contributing to the flux decreases more
slowly with increasing energy than in the absence of magnetic
fields. For a strongly magnetized observer the observed spectrum 
appears to be less modified with respect to the injection spectrum.
This tendency is confirmed by our Monte Carlo simulations,
see Fig.~\ref{fig16}.

By comparing auto-correlations and clustering in scenarios with
and without EGMF we also confirmed that magnetic lensing~\cite{hmrs} is
insignificant for these highly structured EGMF, even for cases
where the observer is immersed in strong magnetic fields.

\section{Discussion and Conclusions} \label{disc.sec}

We studied the effects of non uniform UHECR source distributions
and EGMF imprinted in the multi-pole
moments, auto-correlation function, and cluster statistics
of UHECRs with energies above $10^{19}\,$eV. We compared
scenarios where the sources are distributed homogeneously or
according to the baryon density distribution obtained from
a simulation of the large scale structure of the Universe.
We also compared the case in which the sources are all identical 
and in which their power and spectra of injected UHECRs are distributed
according to functions that maximize cosmic variance. 
The changes in the results due to
variation in the locations and properties of the sources were
evaluated as cosmic variance. The source luminosity function
was chosen close to the critical case where all luminosities
contribute comparably to the observed flux in a logarithmic distribution,
which tends to maximize cosmic variance. In this case we found that
cosmic variance of large scale multi-poles
tends to be larger than the statistical fluctuations if the number
of observed events is $\gtrsim100$. In all other cases, notably
for the auto-correlations and cluster multiplicities, cosmic
variance is small compared to statistical fluctuations for
$N_{\rm obs}\lesssim100$ events.

The influence of EGMF is assessed by comparing 
the case of UHECR propagation when (a) EGMF are negligible and
(b) they are not and are modeled through a simulation of large
scale structure formation and magnetic field evolution. In the latter 
case the magnetic fields were normalized so as to reproduce 
published values of Faraday rotation measures for 
clusters of galaxies and filametns~\cite{bo_review,ryu}.
We chose two cases for the location of the observer: 
in a relatively high field region with $B\simeq0.1\mu$G and
in a negligible field region with $B\simeq10^{-11}\,$G.

In Ref.~\cite{sme} it was found that in the case of structured
magnetic fields an observer immersed in $\sim0.1\,\mu$G fields was
marginally favored whereas isotropy observed at $10^{19}\,$eV
could not be explained by the local component alone. In Ref.~\cite{letter}
and in the present paper we took into account the contribution of sources
beyond the large scale structure simulation box by making use of
the periodic boundary conditions of the simulation box.
As a result we found that the AGASA data is consistent with and even
marginally favors scenarios with magnetic fields
up to a few micro Gauss in galaxy clusters whose structure is obtained
from the large scale structure simulation, and where the sources
follow the baryon distribution. Contrary to Ref.~\cite{sme}
this scenario requires that the observer is in a relatively low field
region, $B\ll0.1\,\mu$G. This also allows to explain isotropy observed
at $10^{19}\,$eV reasonably well. That the fit quality there is
significantly worse than at higher energies, as can be seen from
Tab.~\ref{tab1}, may have to do with significant evolution of the
characteristics of sources at large distances whose contribution
increases with decreasing UHECR energy. The best fit scenario also
predicts a pronounced GZK cutoff as well as considerable
deflection of order $20^\circ$ up to $\simeq10^{20}\,$eV.
{\it Thus, if this scenario is further confirmed by future experiments,
charged particle astronomy may not be possible.}

Structured sources and magnetic fields up to a few micro Gauss in galaxy
clusters and the Earth immersed in relatively low EGMF
$B\ll0.1\,\mu$G in fact seem to be the most realistic
scenario. Thus, already existing UHECR data allow probing the large
scale structure distribution of UHECR sources and magnetic field! 
We have demonstrated how
future detectors such as the Pierre Auger and EUSO projects can
further probe EGMF and UHECR source characteristics: Strongly
magnetized observers predict considerable large scale anisotropies
between $10^{19}\,$eV and $10^{20}\,$eV which is already ruled out by
current data on the percent level. Furthermore, strong fields
surrounding the observer would predict a GZK cutoff that is less
pronounced than for negligible fields. Low auto-correlations at
degree scales imply magnetized sources quite independent of
other source characteristics such as their density. The latter can
only be estimated from the auto-correlations halfway reliably
if magnetic fields have negligible impact on propagation.
The multi-poles for $l\lesssim10$ are relatively insensitive to the
source density and on average tend to be smaller if sources are
magnetized. At the same time, magnetized sources also tend to reduce
cosmic variance of these multi-poles, and as a result their predictions
above $4\times10^{19}\,$eV for future experiments tend to deviate
from isotropy more significantly than for unmagnetized sources.
Unfortunately, especially if source luminosities fluctuate
considerably, it may be difficult to distinguish between structured
and homogeneous source distributions even with next generation experiments.

For the required average local source number density and continuous
power per source above $10^{19}\,$eV we find
$n_s\gtrsim10^{-5}-10^{-4}\,{\rm Mpc}^{-3}$, and
$Q_s\lesssim10^{42}{\rm erg}\,{\rm s}^{-1}$ respectively,
the latter within about one order of magnitude
uncertainty to both sides.  This corresponds to an average UHECR
emissivity of $q_{\rm UHECR} =n_s\,Q_s\sim1.5\times10^{37}
{\rm erg\,Mpc^{-3}\,s^{-1}}$, with an uncertainty likely somewhat
smaller than for the above quantities, since it is fixed by the
observed UHECR flux. Note that the uncertainty in $n_s$ is increased
by the structured EGMF which tends to mix contributions from individual
sources residing in these fields.

A simple study of deflection angles in the context of a constrained
large scale structure simulation has recently been undertaken in
Ref.~\cite{dolag}. They find maximal deflection angles of a few degrees
above $4\times10^{19}\,$eV, considerably smaller
than in our present study. This may be due to at least two reasons:
First, these authors do not study structured sources which tend to
be in regions of high density and magnetic fields which introduces
a bias towards small deflection in the case of Ref.~\cite{dolag}.
In fact, if we take the EGMF
scenario corresponding to scenario 2 in Tab.~\ref{tab1}, but
with a homogeneous source distribution instead, we obtain
average deflection angles of $\simeq61^\circ$ above $4\times10^{19}\,$eV,
$\simeq33^\circ$ above $10^{20}\,$eV, and $\simeq10^\circ$ above
$2\times10^{20}\,$eV. This is smaller than the deflection angles
obtained in scenarios where source positions and strong magnetic fields
are correlated, see Fig.~\ref{fig5}, but still considerably larger than
values obtained in Ref.~\cite{dolag}. In fact, even if the magnetic field
strength is reduced by a factor 10 in our simulations,
the average deflection angle above $4\times10^{19}\,$eV is still
$\sim30^\circ$, only a factor $\simeq2.2$ smaller. This non-linear
behavior of deflection with field normalization is mostly due to the
strongly non-homogeneous character of the EGMF. Note that a field
strength normalization reduction by a factor 10 corresponds to
a left-shift of the x-axis in Fig.~\ref{fig0} of two decades, 
such that the EGMF
is virtually dynamically unimportant in all cells of the simulation.

The remaining discrepancy in typical UHECR deflection between our
present work and Ref.~\cite{dolag}, for cases meaningful to compare,
is most likely due to the different EGMF models used by the different
authors. A detailed consideration of these differences is well beyond
the scope of this paper, although we plan to investigate this issue
further in a forthcoming paper. So little is known about large scale
magnetic fields and their evolution. This makes signatures for
magnetized sources, as discussed in the present paper, even more
important.

\section*{Acknowledgments}
We would like to thank Martin Lemoine and Claudia Isola for
earlier collaborations on the codes partly used in this work.
The work by FM was partially supported by the Research and
Training Network ``The Physics of the Intergalactic Medium''
set up by the European Community
under the contract HPRN-CT2000-00126 RG29185.
GS thanks the Max-Planck-Institut f\"ur Physik, Werner Heisenberg
Institut where part of this work has been performed for
hospitality and financial support.


\begin{thebibliography}{99}

\bibitem{reviews} for recent reviews see J.~W.~Cronin, Rev.~Mod.~Phys.
71 (1999) S165; M.~Nagano, A.~A.~Watson, Rev.~Mod.~Phys. 72 (2000) 689;
A.~V.~Olinto, Phys.~Rept. 333-334 (2000) 329; X.~Bertou,
M.~Boratav, and A.~Letessier-Selvon, Int.~J.~Mod.~Phys. A15 (2000) 2181;
G.~Sigl, Science 291 (2001) 73; F.~Stecker, J. Phys. G. 29 (2003) R47.

\bibitem{school} ``Physics and Astrophysics of Ultra High Energy Cosmic Rays'',
Lecture Notes in Physics, vol.~576 (Springer Verlag, 2001),
eds. M.~Lemoine, G.~Sigl.

\bibitem{sme} G.~Sigl, F.~Miniati, and T.~En\ss lin, Phys.~Rev.~D 68
(2003) 043002.

\bibitem{bm} W.~S.~Burgett and M.~R.~O'Malley, Phys.~Rev.~D 67
(2003) 092002.

\bibitem{uchihori}
Y.~Uchihori, M.~Nagano, M.~Takeda, M.~Teshima, J.~Lloyd-Evans,
and A.~A.~Watson, Astropart.~Phys. 13 (2000) 151.

\bibitem{gzk} K.~Greisen, Phys.~Rev.~Lett. 16 (1966)
748; G.~T.~Zatsepin and V.~A.~Kuzmin, Pis'ma
Zh. Eksp. Teor. Fiz. 4 (1966) 114 [JETP. Lett. 4 (1966) 78].

\bibitem{stecker} F.~W.~Stecker, Phys.~Rev.~Lett. 21 (1968) 1016.

\bibitem{bergman} D.~R.~Bergman, Nucl.~Phys.~B, Proc.~Suppl. 117 (2003)
106; Proc. 28th International Cosmic Ray Conference, Tsukuba, vo.~1,
p.~397 (2003), see
{\sf http://www-rccn.icrr.u-tokyo.ac.jp/icrc2003/PROCEEDINGS/PDF/100.pdf}.

\bibitem{hires} R.~U.~Abasi et al. (HiRes collaboration),
Phys.~Rev.~Lett. 92 (2004) 151101; e-print astro-ph/0208301.

\bibitem{agasa} M.~Takeda et al., Phys. Rev. Lett. 81 (1998) 1163;
Astrophys. J. 522 (1999) 225; Hayashida et al.,
e-print astro-ph/0008102; see also
{\sf http~://www-akeno.icrr.u-tokyo.ac.jp/AGASA/}.

\bibitem{auger} J.~W.~Cronin, Nucl.~Phys.~B (Proc.~Suppl.) 28B (1992)
213; The Pierre Auger Observatory Design Report (ed.~2), March 1997;
see also {\sf http://www.auger.org}.

\bibitem{slb} G.~Sigl, M.~Lemoine, and P.~Biermann,
Astropart.~Phys. 10 (1999) 141.

\bibitem{ils} C.~Isola, M.~Lemoine, and G.~Sigl, Phys.~Rev.~D 65 (2002) 023004.

\bibitem{lsb}
M.~Lemoine, G.~Sigl, and P.~Biermann, e-print astro-ph/9903124.

\bibitem{sse} T.~Stanev, e-print astro-ph/0303123; T.~Stanev, D.~Seckel,
and R.~Engel, Phys.~Rev.~D 68 (2003) 103004;
see also T.~Stanev, R.~Engel, A.~Mucke, R.~J.~Protheroe, and J.~P.~Rachen,
Phys.~Rev.~D 62 (2000) 093005.

\bibitem{is} C.~Isola and G.~Sigl, Phys.~Rev.~D 66 (2002) 083002.

\bibitem{dolag}
K.~Dolag, D.~Grasso, V.~Springel, and I.~Tkachev, e-print astro-ph/0310902.

\bibitem{sommers}
P.~Sommers, Astropart.~Phys. 14 (2001) 271.

\bibitem{bdm} P.~Blasi and D.~De Marco, Astropart.~Phys. 20 (2004) 559.

\bibitem{ynts} H.~Yoshiguchi, S.~Nagataki, S.~Tsubaki, and K.~Sato,
Astrophys.~J. 586 (2003) 1211; H.~Yoshiguchi, S.~Nagataki, and K.~Sato,
Astrophys.~J. 592 (2003) 311; Astrophys.J. 596 (2003) 1044.

\bibitem{tanco}
G.~Medina Tanco, ``Cosmic magnetic fields from the perspective
of ultra-high-energy cosmic rays propagation'', Lect.~Notes~Phys.
576 (2001) 155.

\bibitem{letter} G.~Sigl, F.~Miniati, and T.~En\ss lin, e-print
astro-ph/0309695.

\bibitem{euso} See {\sf http://www.euso-mission.org}.

\bibitem{medina} G.~Medina-Tanco, E.~M.~De~Gouveia~Dal~Pino,
and J.~E.~Horvath, e-print astro-ph/9707041.

\bibitem{miniati} F.~Miniati, Mon.~Not.~R.~Astron.~Soc. 337 (2002) 199.

\bibitem{ryu} D.~Ryu, H.~Kang, and P.~L.~Biermann,
Astron.~Astrophys. 335 (1998) 19.

\bibitem{kcor97} R. M. Kulsrud, R. Cen, J. P. Ostriker, and D. Ryu,
Astrophys.~J., 480 (1997) 481.
 
\bibitem{kronberg99} P.~P.~Kronberg, H.~Lesch, and H.~Ulrich,
Astrophys.~J. 511 (1999) 56.

\bibitem{bo_review} P.~P.~Kronberg, Reports of Progress in
Physics 58 (1994) 325; J.~P.~Vall{\'e}e, Fundamentals of Cosmic
Physics, Vol.~19 (1997) 1; T.~E.~Clarke, P.~P.~Kronberg, and
H.~{B{\" o}hringer}, Astrophys.~J.~Lett. 547 (2001) L111; J.-L.~Han
and R.~Wielebinski, Chinese Journal of Astronomy and Astrophysics 2
(2002) 293 [e-print astro-ph/0209090];
P.~P.~Kronberg, Physics Today 55, December 2002, p.~40.

\bibitem{minthe} F.~Miniati, Ph.D. Thesis, University of Minnesota 2000;
F.~Miniati, T.~W.~Jones, H.~Kang, and D.~Ryu, Astrophys.~J. 562 (2001) 233.

\bibitem{fufe99} R.~Fusco-Femiano, D.~ Dal~Fiume, L.~Feretti, G.~Giovannini,
P. Grandi, G.~Matt, S.~Molendi, A.~Santangelo, 513 (1999) L21.

\bibitem{bagchi02} J. Bagchi, T. A. Ensslin, F. Miniati, C.S. Stalin,
M. Singh, S. Raychaudhury, N.B. Humeshkar, New Astronomy, 7 (2002) 249.

\bibitem{cm} J.~Chiang and R.~Mukherjee, Astrophys.~J. 496 (1998) 752.

\bibitem{bs-rev} P.~Bhattacharjee and G.~Sigl,
Phys.~Rept. 327 (2000) 109; L.~Anchordoqui, T.~Paul, S.~Reucroft,
and J.~Swain, Int.~J.~Mod.~Phys. A18 (2003) 2229.

\bibitem{sugar} R.~G.~Brownlee et al., Can.~J.~Phys. 46 (1968) S259;
M.~M.~Winn et al., J.~Phys.~G 12 (1986) 653; R.~W.~Clay et al.,
Astron.~Astrophys. 255 (1992) 167; L.~J.~Kewley, R.~W.~Clay, and
B.~R.~Dawson, Astropart.~Phys. 5 (1996) 69; C.~J.~Bell et al.,
J.~Phys.~A 7 (1974) 990; see also
{\sf http://www.physics.usyd.edu.au/hienergy/sugar.html}.

\bibitem{k-s} M.~Kachelrie\ss\ and D.~V.~Semikoz, Phys.~Lett. B577 (2003) 1;
H.~B.~Kim and P.~Tinyakov, e-print astro-ph/0306413.

\bibitem{sjm} B.~T.~Stokes, C.~C.~H.~Jui, and J.~N.~Matthews,
Astropart.~Phys. 21 (2004) 95.

\bibitem{anchor_iso} L.~Anchordoqui et al., Phys.~Rev.~D 68 (2003) 083004.

\bibitem{wm} E.~Waxman and J.~Miralda-Escud\'{e}:
Astrophys.~J.472 (1996) L89.

\bibitem{hmrs} see, e.g., D.~Harari, S.~Mollerach, E.~Roulet, F.~Sanchez,
JHEP 0203 (2002) 045.

\end{thebibliography}
\end{document}